\documentclass[prd,superscriptaddress,amsfonts,amssymb,,twocolumn,amsmath,showpacs]{revtex4-2}
\usepackage{bm}
\usepackage{amsfonts}
\usepackage{latexsym}
\usepackage[latin1]{inputenc}
\usepackage{graphicx}
\usepackage{amsmath}
\usepackage{palatino}
\usepackage{mathpazo}
\usepackage{textcomp}
\usepackage{float}
\usepackage{booktabs}
\usepackage{dcolumn}
\usepackage{booktabs}
\usepackage{multirow}
\usepackage{hyperref}
\hypersetup{colorlinks,citecolor=red}
\usepackage{amsmath}
\usepackage{xcolor}
\usepackage{orcidlink}
\usepackage[caption=false]{subfig}
\usepackage{commath}
\def\jnl@style{\it}
\def\aaref@jnl#1{{\jnl@style#1}}

\def\aaref@jnl#1{{\jnl@style#1}}

\def\aj{\aaref@jnl{AJ}}                   % Astronomical Journal
\def\apj{\aaref@jnl{ApJ}}                 % Astrophysical Journal
\def\apjl{\aaref@jnl{ApJ}}                % Astrophysical Journal, Letters
\def\apjs{\aaref@jnl{ApJS}}               % Astrophysical Journal, Supplement
\def\apss{\aaref@jnl{Ap\&SS}}             % Astrophysics and Space Science
\def\aap{\aaref@jnl{A\&A}}                % Astronomy and Astrophysics
\def\aapr{\aaref@jnl{A\&A~Rev.}}          % Astronomy and Astrophysics Reviews
\def\aaps{\aaref@jnl{A\&AS}}              % Astronomy and Astrophysics, Supplement
\def\mnras{\aaref@jnl{Mon.~Not.~Roy.~Astron.~Soc.}}             % Monthly Notices of the RAS
\def\prd{\aaref@jnl{Phys.~Rev.~D}}        % Physical Review D
\def\prc{\aaref@jnl{Phys.~Rev.~C}}  % Physical Review C
\def\prl{\aaref@jnl{Phys.~Rev.~Lett.}}    % Physical Review Letters
\def\qjras{\aaref@jnl{QJRAS}}             % Quarterly Journal of the RAS
\def\skytel{\aaref@jnl{S\&T}}             % Sky and Telescope
\def\ssr{\aaref@jnl{Space~Sci.~Rev.}}     % Space Science Reviews
\def\zap{\aaref@jnl{ZAp}}                 % Zeitschrift fuer Astrophysik
\def\nat{\aaref@jnl{Nature}}              % Nature
\def\aplett{\aaref@jnl{Astrophys.~Lett.}} % Astrophysics Letters
\def\apspr{\aaref@jnl{Astrophys.~Space~Phys.~Res.}} % Astrophysics Space Physics Research
\def\physrep{\aaref@jnl{Phys.~Rep.}}      % Physics Reports
\def\physscr{\aaref@jnl{Phys.~Scr}}       % Physica Scripta
\def\commat{\aaref@jnl{Comm.~Math.~Phys.}}              % Communications in Mathematical Physics
\def\science{\aaref@jnl{Science}}               % Science
\def\cqg{\aaref@jnl{Classical Quant.~Grav.}}            % Classical and Quantum Gravity
\def\jpcs{\aaref@jnl{JPCS}}                                     % Journal of Physics Conference Series
\def\ijmpd{\aaref@jnl{Int.~J.~Mod.~Phys.~D}}                    % International Journal of Modern Physics D
\def\grg{\aaref@jnl{Gen.~Relat.~Gravit.}}               % General Relativity and Gravitation
\def\rpp{\aaref@jnl{Rep.~Prog.~Phys.}}          % Reports on Progress in Physics
\def\npa{\aaref@jnl{Nucl.~Phys.~A}}        % Nuclear Physics A
\def\lrr{\aaref@jnl{Living Rev.~Rel.}}                   % Living reviews in relativity
\def\jcap{\aaref@jnl{J.~Cosmology Astropart.~Phys.}}    % Journal of cosmology and astroparticle physics
\def\rmp{\aaref@jnl{Rev.~Mod.~Phys.}}   %Reviews of modern physics
\def\epjc{\aaref@jnl{Eur.~Phys.~J.~C}}

\newcommand{\udt}[3]{#1^{#2}_{\phantom{#2}#3}}

\allowdisplaybreaks[1]
\begin{document}

\color{black}       %% For one column
%
%\title{\bf Teleparallel scalar field reconstruction of matter bounce scenario}
\title{\bf Non-minimally coupled teleparallel scalar field reconstruction of matter bounce scenario}

\author{S. K. Tripathy \orcidlink{0000-0001-5154-2297}}
\email{tripathy\_sunil@rediffmail.com}
\affiliation{Department of Physics, Indira Gandhi Institute of Technology, Sarang, Dhenkanal, Odisha-759146, India.}

\author{Sasmita Pal \orcidlink{0009-0005-0812-6059}}
\email{palsasmita57@gmail.com}
\affiliation{Department of Physics, Indira Gandhi Institute of Technology, Sarang, Dhenkanal, Odisha-759146, India.}

\author{B. Mishra\orcidlink{0000-0001-5527-3565} }
\email{bivu@hyderabad.bits-pilani.ac.in }
\affiliation{Department of Mathematics, Birla Institute of Technology and Science-Pilani, Hyderabad Campus, Hyderabad-500078, India.}

%\date{\today}
\begin{abstract}
Teleparallel description of gravity theories where the gravity is mediated through the tetrad field and consequent torsion provide an alternative route to explain the late time cosmic speed up issue. Generalization of the teleparallel gravity theory with different functional forms of the torsion scalar $T$ leads to $f(T)$ gravity. The role of scalar field played in addressing issues in cosmology and astrophysics has developed an interest in the inclusion of a scalar field along with an interaction potential in the action. Such a generalized gravity theory is dubbed as $f(T,\phi)$ theory. We have explored such a gravity theory to reconstruct the interaction potential of the scalar field required for an extended matter bounce scenario. The cosmological implications of the reconstructed scalar field potential are studied  considering two viable and well known functional forms of $f(T,\phi)$. The energy conditions of these model are discussed to assess the viability of the cosmological models.
\end{abstract}

\maketitle

\section{Introduction} \label{SEC I}
Teleparallel equivalent of General Relativity (TEGR) is an alternative geometrical approach to introduce an analogous description of gravity.  The teleparallel theory of gravity is based on teleparallel geometry, where the gravitational source comes from a torsional formulation rather than the curvature scalar of GR \cite{Cai:2015emx}. In teleparallel theory, the fundamental objects are tetrads, which are orthogonal fields based on the four-dimensional spacetime tangent space \cite{Hohmann_2018_97_teleparallel,farrugia2020gravitoelectromagnetism,Aldrovandi:2013wha}. Initially, it was believed that the additional six degrees of freedom of this structure could unify electromagnetism and gravity. However, later this idea was found to be incorrect, and led Einstein to propose an equivalent theory to GR known as TEGR. \cite{Krssak:2018ywd,Bahamonde:2015zma,Dialektopoulos:2021ryi,Aldrovandi:2013wha}. In fact, the additional six degrees of freedom involved in TEGR are related to the local Lorentz invariance of the theory rather than electromagnetism \cite{Ferraro:2006jd}. One popular and interesting model beyond TEGR is the $f(T)$ theory of gravity \cite{Basilakos:2018arq,Ferraro:2006jd, Cai:2015emx,Bengochea:2008gz}, which involves the replacement of torsion scalar $T$ in teleparallel gravity (TG) by a generic function of the torsion scalar $f(T)$. This modified gravity theory is different from the simplified geometrical $f(R)$ gravity in the sense that $f(T)$ theory results in second-order field equations, whereas $f(R)$ gravity involves fourth-order field equations \cite{Basilakos:2011rx,Basilakos:2013rua,Hohmann:2018rwf}. Further, $f(T)$ gravity can be used to explain the current accelerated expansion of the Universe without introducing any dark energy (DE) concept\cite{DUCHANIYA2024f(T),Farrugia:2016qqe}. This generic extension of TEGR, i.e. $f(T)$ theory occupies an important position in the context of different investigations in cosmology. It provides useful insights into the issue of the accelerating cosmic expansion of the Universe and constructs cosmological models mostly based on observational data \cite{MNRASf(T)data}. Also, this theory has been successfully used to analyze different epochs of evolution of Universe using robust techniques like the dynamical system analysis \cite{Hohmann:2017}. The viability with commendable success in studying the four solar system tests of gravity, namely the perihelion precession, light bending, Shapiro time delay, and gravitational redshift in the $f (T)$ gravity formalism has developed a lot of research interests in recent times\cite{Farrugia:2016xcw}. \\

Of late, an alternative formulation with a coupling occurring between the scalar field $\phi$ and the torsion scalar $T$ of the form of $\xi T \phi^2$ has gained some research attractions\cite{Geng:2011aj,Kadam:2022lgq}. This approach leads to an interesting phenomenon, such as phantom behavior and the dynamic crossing of the phantom barrier. The dynamics of such models have been analyzed through dynamical systems along with  the observational constraints put on some cosmological parameters \cite{Geng:2011aj, Geng_2012}. In this paper, we consider the non minimal coupling to the torsion scalar $T$, which takes a different approach, and examine a non minimal coupling between the scalar field and the general function of the torsion scalar $T$. The action of the proposed model contains the general form of the non-minimally coupled scalar field form $F(\phi)G(T)$. These generalizations are assumed to be an extension of $f(T)$ gravity in terms of the scalar field and is termed as $f(T,\phi)$ gravity \cite{Gonzalez-Espinoza:2020jss, Duchaniya_2023nor}. Substantial works are available in the literature that highlight the validity of the $f(T, \phi)$ gravity theory such as  the dynamical system analysis \cite{Gonzalez-Espinoza:2020jss}, the stability analysis using the scalar perturbations to study the viability conditions for the absence of ghost, gradient and tachyonic
instabilities in the presence of a general barotropic perfect fluid \cite{Gonzalez-Espinoza:2021mwr}. Also, the reconstruction of power law function of the torsion scalar is carried out in detail in Ref. \cite{Gonzalezreconstruction2021}.\\

 In inflationary cosmology, modes well inside the Hubble radius are assumed to be initially in the adiabatic vacuum to have a nearly scale invariant spectrum \cite{Haro_2014}. Inflation is believed to be started at the energies of Planck scale. However observations for the power spectrum for scalar perturbations for the modes that exit the Hubble radius hinted for a classical Universe with the modes evolving prior to the inflationary scenario and at energies much lower than the Planck scale. The bouncing scenario is an assumed phenomenon in which the Universe follows certain specific evolutionary process quite different than the standard cosmological predictions. Initially the Universe is believed to undergo a matter dominated contraction phase followed by a non-singular bounce and then there is a casual generation for fluctuation leading to expansion. Accordingly one can visualise the connection with the Hubble parameter $H=\frac{\dot{a}}{a}$, $a$ being the scale factor. For a bouncing scenario, the Hubble parameter passes through zero $(H=0)$ at the bouncing epoch from negative values $(H<0)$ in the pre-bounce contraction phase to positive values $(H>0)$ in the expansion phase. In the matter bounce scenario, it is believed that quantum fluctuations from vacuum perturbations exiting the Hubble radius during a matter dominated contracting phase lead to a scale-invariant spectrum \cite{CAI201125}. This model solves the horizon problem and improves the flatness issue appearing in GR.

 The semi-classical framework of the Loop Quantum Cosmology (LQC) obtained from Loop Quantum Gravity (LQG) resolves the issue of singularities appearing in GR and predicts a bouncing Universe. In LQC, the bounce occurs when the space-time singularity is of the order of the Planck scale \cite{Ashtekar_2011}. Essentially, LQC assumes a discrete Hilbert space where the quantum states are represented by almost periodic functions of the dynamical part of the connection \cite{Ashtekar_2011, deHaro2015} which requires a well defined Hamiltonian. The holonomy corrected LQG, where the holonomy and flux operators are used to quantize the Hamiltonian constraint provide matter bounce solutions that are more consistent with the experimental data \cite{EdwardWilson-Ewing_2013, PhysRevD.90.124083}. 

An overview of the categorisations of torsion tensors and discussion regarding the representation of gravitational waves in scalar torsion gravity is presented in \cite{Cai:2015emx}. In the minimal case, the scalar torsion gravity is entirely equivalent to standard quintessence. At the same time, in the nonminimal scenario, it exhibits a more complex structure, displaying behaviour similar to quintessence or phantom, including crossing the phantom-divide \cite{Geng:2011aj}. The interesting and well-studied models in scalar-tensor theories with their field equations and how they transform under conformal transformations have been reviewed in \cite{bahamonde:2021teleparallel}.
Bouncing cosmology study is available in the literature in different modified gravity theories \cite{Agrawal_A_2023_matter,Agrawal_2021,Agrawal_2024,NOJIRI2022100984,Oikonomou_2015, MISHRA2024138968,ELIZALDE2020114984}. We are motivated here to reconstruct the scalar field potential required to witness an extended matter bounce scenario within a setup of $f(T,\phi)$ gravity. \\

The paper is organised as follows, in Sec.~\ref{SECII}, we present the basic equations of  $f(T,\phi)$ gravity in an isotropic an homogeneous space time. In the $f(T,\phi)$ gravity, the action contains the contribution from a scalar field along with an interaction potential. In Sec.~\ref{SECIII}, we integrate the Klein-Gordon equation to reconstruct the scalar field potential for an extended matter bounce scenario motivated from LQC. The cosmological implications of the reconstructed scalar field potential is studied for two different functions of the torsion scalar in Sec.~\ref{SECIV}. The energy conditions of the cosmological models witnessing an extended matter bounce scenario are discussed in Sec.~\ref{SECV}. In Sec.~\ref{SECVI}, the discussions and conclusions of the models are presented.

\section{Field Equations of \texorpdfstring{$f(T,\phi)$}{} gravity }\label{SECII}

The action in the $f(T,\phi)$ gravity theory may be written as \cite{Gonzalez-Espinoza:2021mwr, Duchaniya_2023}
\begin{equation}\label{1}
    S =\int d^{4}x~~e[f(T,\phi)+X]+S_{m}+S_{r}\,,
\end{equation}
where $f(T,\phi)$ is an arbitrary function of a scalar field $\phi$ with kinetic energy $X= -\partial_{\mu} \phi \partial^{\mu} \phi/2$. $T$ is the torsion scalar $T = S_{A}^{\, \, \, \, \, \mu \nu} T^{A}_{\, \, \, \, \, \mu \nu }$ where $S_{A}^{\, \, \, \, \, \mu \nu}$ and $T^{A}_{\, \, \, \, \, \mu \nu }$ respectively represent the superpotential and the torsion tensor. $e = \det[e^A_{\, \, \, \mu}] = \sqrt{-g}$ is the determinant of the tetrad field $e^{A}_{\, \, \, \mu}$ with $A = 0,1,2,3$. The tetrad field satisfies the orthogonality condition, $e^{B}_{\, \, \, \mu} e_{A}^{\, \, \, \mu}=\delta_A^B$. The Minkowski tangent space metric $\eta_{AB}=\text{diag}(-1,1,1,1)$  and the metric tensor $g_{\mu \nu}$ are connected through the relation $g_{\mu \nu}=\eta_{AB} e^{A}_{\, \, \, \mu} e^{B}_{\, \, \, \nu}$, where $\eta_{AB}=(-1,1,1,1)$. The spin connection is denoted by $\udt{\omega}{A}{B\mu}$. $S_{m}$ and $S_{r}$ respectively denote the matter action and the action for radiation. In the action \eqref{1}, we have chosen the units so that $8\pi G=c=1$, where $G$ is the Newtonian Gravitational constant and $c$ is the speed of light in vacuum. Moreover, the Latin indices represent the Minkowski space, while Greek indices  correspond to the general manifold.

The torsion tensor is given by

\begin{equation}\label{2}
    T^{A}_{\, \, \, \mu \nu} = \partial_{\mu} e^{A}_{\, \, \, \nu}-\partial_{\nu}e^{A}_{\, \, \, \mu}+\omega^{A}_{\, \, \, B\mu}e^{B}_{\, \, \, \nu} - \omega^{A}_{\, \, \, B\nu}e^{B}_{\, \, \, \mu},
\end{equation}

so that we can define the contortion tensor as

\begin{equation}\label{3}
    K^{\rho}_{~~\nu \mu}= \frac{1}{2}(T^{~~\rho}_{\nu~~~\mu}+T^{~~\rho}_{\mu~~~\nu}-T^{\rho}_{~~~\nu\mu}).
\end{equation}
In terms of the contortion tensor, the superpotential is expressed as,
\begin{equation} \label{4}
    S_{\rho}^{~~\mu \nu}=\frac{1}{2}(K^{\mu \nu}_{~~~\rho}+\delta^{\mu}_{\rho}T^{\theta \nu}_{~~~\theta}-\delta^{\nu}_{\rho}T^{\theta \mu}_{~~~\theta})\,.
\end{equation} 
In the framework of teleparallel gravity, the Weitzenb$\ddot{o}$ck gauge can be understood as selecting a frame of reference in which inertial effects, such as fictitious forces caused by acceleration or rotation, can be eliminated. This decision streamlines the equations by reducing complexity by eliminating the spin connection ($\omega^{A}_{\, \, \, B\mu}=0$), thus leaving the tetrad field as the only descriptor of gravitational interaction. 

By varying the action with respect to the tetrad field, we obtain the field equations for the $f(T,\phi)$ gravity  \cite{Hohmann:2018rwf, Duchaniya_2023}.  Also, we have the torsion scalar as

\begin{equation}\label{5}
    T = -R + {2}e^{-1}\partial_{\mu}(eT^{\alpha\mu}_{~~\alpha})\,.
\end{equation}
In this paper, we choose the tetrad of the form,  $e^{A}_{\, \, \, \mu}= \text{diag}\left(1, a(t), a(t), a(t)\right)$. This choice is a proper tetrad i.e. one which has a vanishing spin connection as it satisfies the antisymmetric field equations of $f(T,\phi)$ gravity.
In a background of a homogeneous and isotropic flat Friedmann-Lema\^{i}tre-Robertson-Walker (FLRW) space time 
\begin{equation}\label{6}
    ds^{2}=-dt^{2}+a^{2}(t)[dx^2+dy^2+dz^2],
\end{equation}
the field equations in the framework of $f(T,\phi)$ cosmology become

\begin{eqnarray}
    f(T,\phi)-X-2Tf_{T}&=&\rho_{m}+\rho_{r}, \label{7}\\
    f(T,\phi)+X-2Tf_{T}-4\dot{H}f_{T}-4H\dot{f}_{T} &=& -p_{r},\label{8}\\
    -3H\dot{\phi}-\ddot{\phi}+f_{\phi}&=& 0. \label{9}
\end{eqnarray}

We have $a(t)$ as the scale factor and the  Hubble parameter as $H\equiv\frac{\dot{a}}{a}$ with an over dot denoting the derivative with respect to the cosmic time $t$. $\rho_{m}$, $\rho_{r}$ are the respective energy densities for matter and radiation fields. The matter field is assumed to be composed of pressureless dust and the radiation pressure is denoted by $p_{r}=\frac{1}{3} \rho_r$. In the above equations ~\eqref{7}--\eqref{9}, we adopt the shorthand notations $f\equiv f(T,\phi)$, $f_{T}=\frac{\partial f}{\partial T}$ and $f_{\phi}=\frac{\partial f}{\partial \phi}$. The torsion scalar for the present set up is obtained as $T=6H^{2}$.

For a non-minimal coupling function $f(T,\phi)$ with a scalar field potential $V(\phi)$ 
\begin{equation}\label{10}
    f(T,\phi)=-\frac{T}{2}-G(T)-V(\phi)\,,
\end{equation}
the equivalent Friedmann equations of the $f(T,\phi)$ cosmology are written as

\begin{eqnarray}
     & &\frac{\dot{\phi}^2}{2}+V(\phi)-2TG_{T}+G(T)+\rho_{m}+\rho_{r} = 3H^2,\label{11}\\
    &-&\frac{\dot{\phi}^2}{2}+V(\phi)-2TG_{T}+G(T)\nonumber\\
    &-&4\dot{H}(G_{T}+2TG_{TT})-\frac{1}{3}\rho_{r} = 2\dot{H}+3H^2,\label{12}\\
& &\ddot{\phi}+3H\dot{\phi}+V_{\phi}(\phi) = 0.\label{13}
\end{eqnarray}

$G(T)$ is an arbitrary function of torsion scalar suitably chosen to get viable cosmological models. Also, we consider the notations: $G_{T}\equiv \frac{\partial G(T)}{\partial T}$, $G_{TT} \equiv \frac{\partial^2 G(T)}{\partial T^2}$ and $V_{\phi}=\frac{\partial V(\phi)}{\partial \phi}$.

Comparing the  Eqs.~\eqref{11}--\eqref{12} with the usual Friedmann equations 
\begin{eqnarray}
    3H^{2} &=& \rho_{m}+\rho_{r}+\rho_{d}, \label{14}\\
    2\dot{H}+3H^2 &=& -(p_r+p_d), \label{15}
\end{eqnarray}
the energy density and the pressure of the dark sector are obtained as

\begin{align}
    \rho_{d} &= \frac{\dot{\phi}^2}{2}+V(\phi)-2TG_{T}+G(T), \label{16}\\
    p_{d} &= \frac{\dot{\phi}^2}{2}-V(\phi)+2TG_{T}-G(T)+4\dot{H}(G_{T}+2TG_{TT}). \label{17}
\end{align}

It is now straightforward to write the equation of state (EoS) parameter of the dark sector
\begin{equation}
    \omega_{d}=\frac{p_{d}}{\rho_{d}}=\frac{\frac{\dot{\phi}^2}{2}-V(\phi)+2TG_{T}-G(T)+4\dot{H}(G_{T}+2TG_{TT})}{\frac{\dot{\phi}^2}{2}+V(\phi)-2TG_{T}+G(T)},
\end{equation}
which may also be expressed as,
 \begin{equation}
    \omega_{d}=-1+\frac{\dot{\phi}^2+4\dot{H}(G_{T}+2TG_{TT})}{\frac{\dot{\phi}^2}{2}+V(\phi)-2TG_{T}+G(T)}.
\end{equation}

\section{scalar field potential reconstruction for an extended matter bounce scenario}\label{SECIII}
The scalar field potential plays a vital role in the occurrence of a cosmic bounce. Given a bouncing scenario within the specified gravity theory, it is possible to assess the evolutionary behaviour of the scalar field potential. In the present section, we wish to consider some extended matter bounce scenario and reconstruct the scalar field potential.

The  Klein-Gordon equation \eqref{13} for the scalar field may be expressed as,
\begin{equation}\label{19}
    \frac{d}{dt}\left(\frac{\dot{\phi}^{2}}{2}+V(\phi)\right)=-3H\dot{\phi}^{2}.
\end{equation}

Integrating the  Klein-Gordon equation provides the scalar field potential as
\begin{eqnarray}\label{20}
    V(\phi) = -\frac{\dot{\phi}^{2}}{2}-3\int~dt~H\dot{\phi}^{2}.
\end{eqnarray}

An extended matter bounce scenario may be modelled as
\begin{equation}
    a=a_0~(1+\alpha t^2)^{2/3},\label{22}
\end{equation}
where $a_0$ is the scale factor at the bouncing epoch $t=0$. For the sake of brevity, we take $a_0=1$. The single unknown parameter $\alpha$ in the scale factor may be fixed from the present value of the deceleration parameter. One should note that, the parameter $\alpha$ has a greater role in LQG, being associated with the critical density $\rho_c$. In the holonomy corrected LQC, the Friedman equation is expressed as
\begin{equation}
    H^2=\frac{\rho}{3}\left(1-\frac{\rho}{\rho_c}\right),
\end{equation}
where the critical density $\rho_c=\frac{4\sqrt{3}}{\delta^3}$, $\delta$ being the Barbero-Immirzi parameter \cite{Haro_2014}. The corresponding scale factor may be
\begin{equation}
    a(t)=\left(\frac{3}{4}\rho_c t^2+1\right)^{\sigma}.
\end{equation}
In the matter bounce scenario within the LQG, we have $\sigma=1/3$. For an extended matter bounce scenario we may consider  $\sigma=2/3$. One should note that, the parameter $\alpha$ has a greater role in LQG, being associated with the critical density $\rho_c$ through the relationship $\alpha=\frac{3}{4}\rho_c$. The Hubble parameter and the deceleration parameter for such an extended matter bounce scenario are expressed as
\begin{eqnarray}
    H &=& \frac{4\alpha t} {3(1+\alpha t^2)},\\
    q &=& -\frac{3}{4\alpha t^2}-\frac{1}{4} . 
\end{eqnarray}

Considering the present epoch to be $t_0=1$, the present epoch deceleration parameter becomes $q_0=-\frac{3}{4\alpha} -\frac{1} {4}$, so that the model parameter $\alpha$ is expressed in terms of $q_0$ as $\alpha = -\left(\frac{3}{4q_0+1}\right)$. The deceleration parameter is an important cosmographic parameter and its value at the present epoch has been constrained from different observational data. While  Camarena et al.  have constrained it to be $q_0 = -1.08\pm 0.29$ \cite{Camarena2020}, Goswami et al. from a detailed analysis obtained it as $q_0=-0.59$ \cite{Goswami2021}.

\begin{figure}[H]
\centering
\minipage{0.50\textwidth}
\includegraphics[width=\textwidth]{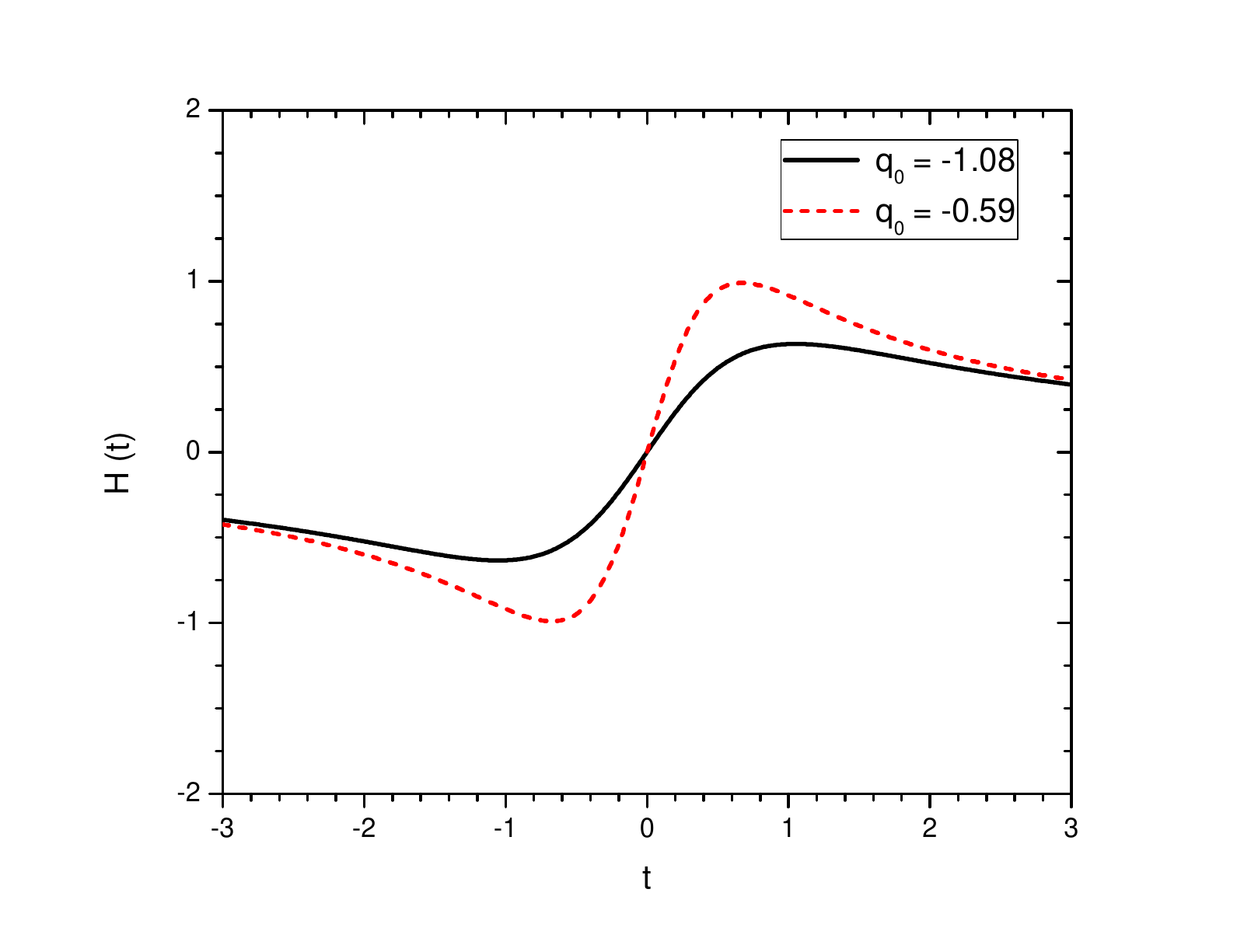}
\endminipage\hfill
\minipage{0.50\textwidth}
\includegraphics[width=\textwidth]{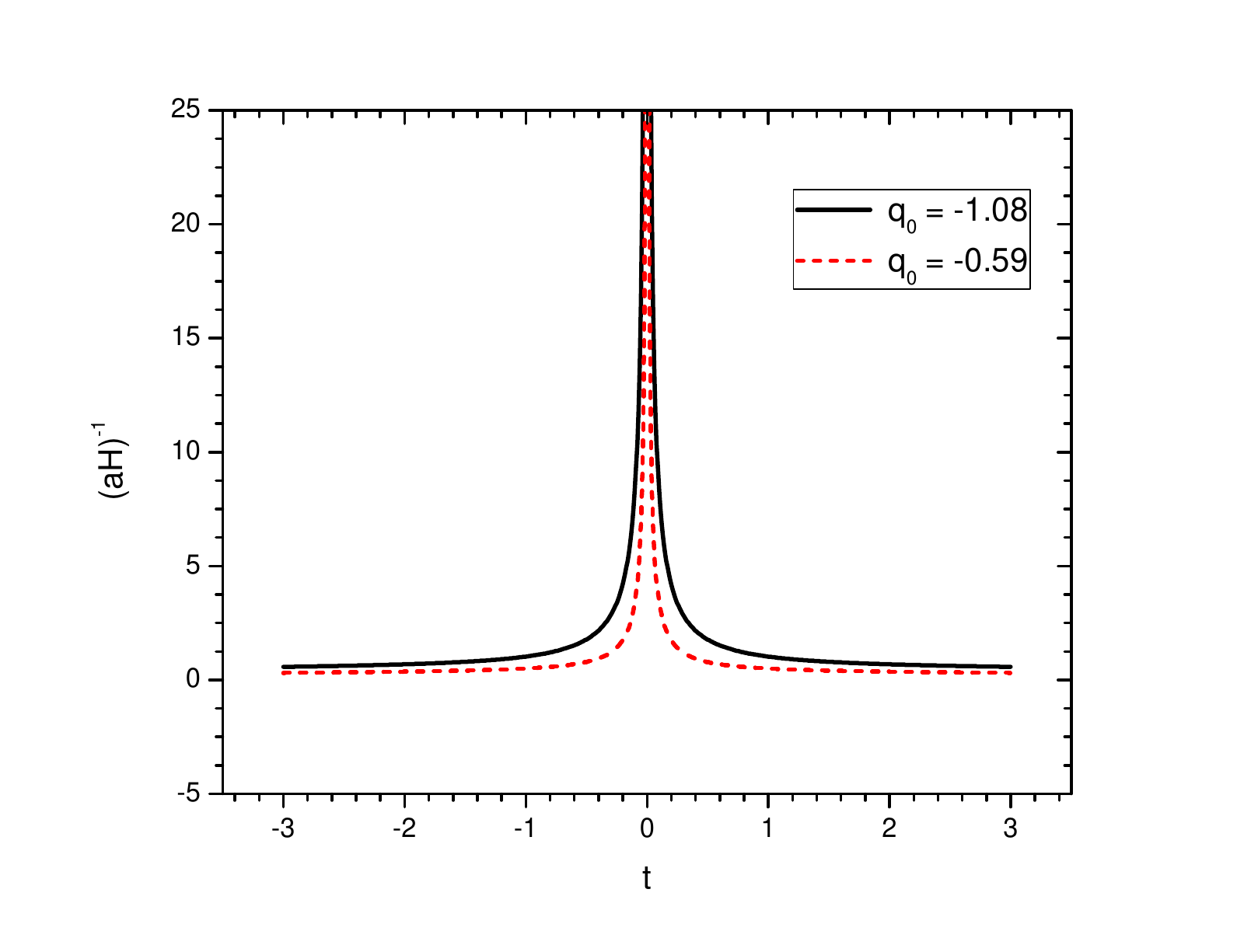}
\endminipage\hfill
\caption{(a) Top panel: Evolution of the Hubble parameter in the extended matter bounce scenario. (b) Bottom panel: Hubble radius.}\label{FIG.1}
\end{figure}

In FIG.\ref{FIG.1}, we show the behaviour of the Hubble parameter (top panel) and the corresponding Hubble radius for the assumed extended matter bounce scenario for two different present epoch values of the deceleration parameter. The Hubble parameter obeys the bouncing conditions and evolves from negative values in the pre-bounce phase to positive values in the post bounce phase of evolution. As usual, at the bouncing epoch, the Hubble parameter vanishes. The magnitude of the Hubble radius (bottom panel) smoothly decreases from high values at the bouncing epoch as we move away from the bounce. Such a behaviour of the Hubble radius is required for  a viable bouncing scenario.  The late-time acceleration of the Universe is ensured by the diminishing trend of the cosmic Hubble radius. The choice of the present epoch value deceleration parameter affects the evolution trajectory of the Hubble parameter and the Hubble radius. 

 We may assume the scalar field to have a power law dependence on the scale factor as $\phi=\phi_0 a^m$ so that $\dot \phi= m \phi H$. The scalar field potential may now be reconstructed from the assumed bouncing scale factor as

       \begin{eqnarray}
           V(t) &=&  \frac{8}{9(2m-3)} \phi_0^2 m^2 ~ \left[6-(2m-3) \alpha t^2 ~\right] (1+\alpha t^2) ^ \frac{4m-6}{3} \nonumber\\
           &-&\frac{32}{3(4m-3)} \phi_0^2 \alpha  (1+\alpha t^2) ^\frac{4m-3}{3}. 
       \end{eqnarray}

We may also express the reconstructed scalar field potential in terms of the scalar field as
\begin{eqnarray}
    V(\phi) &=&  \frac{8}{9(2m-3)} \phi_0^2 m^2 \nonumber\\
    &\times& 
    \left[(3+2m)\left(\frac{\phi}{\phi_0}\right)^{\frac{2m-3}{m}}-(2m-3) \left(\frac{\phi}{\phi_0}\right)^{\frac{4m-3}{2m}} \right] \nonumber\\
           &-&\frac{32}{3(4m-3)} \phi_0^2 \alpha  \left(\frac{\phi}{\phi_0}\right) ^\frac{4m-3}{2m}.
\end{eqnarray}

\begin{figure}[ht!]
\centering
\includegraphics[scale=0.38]{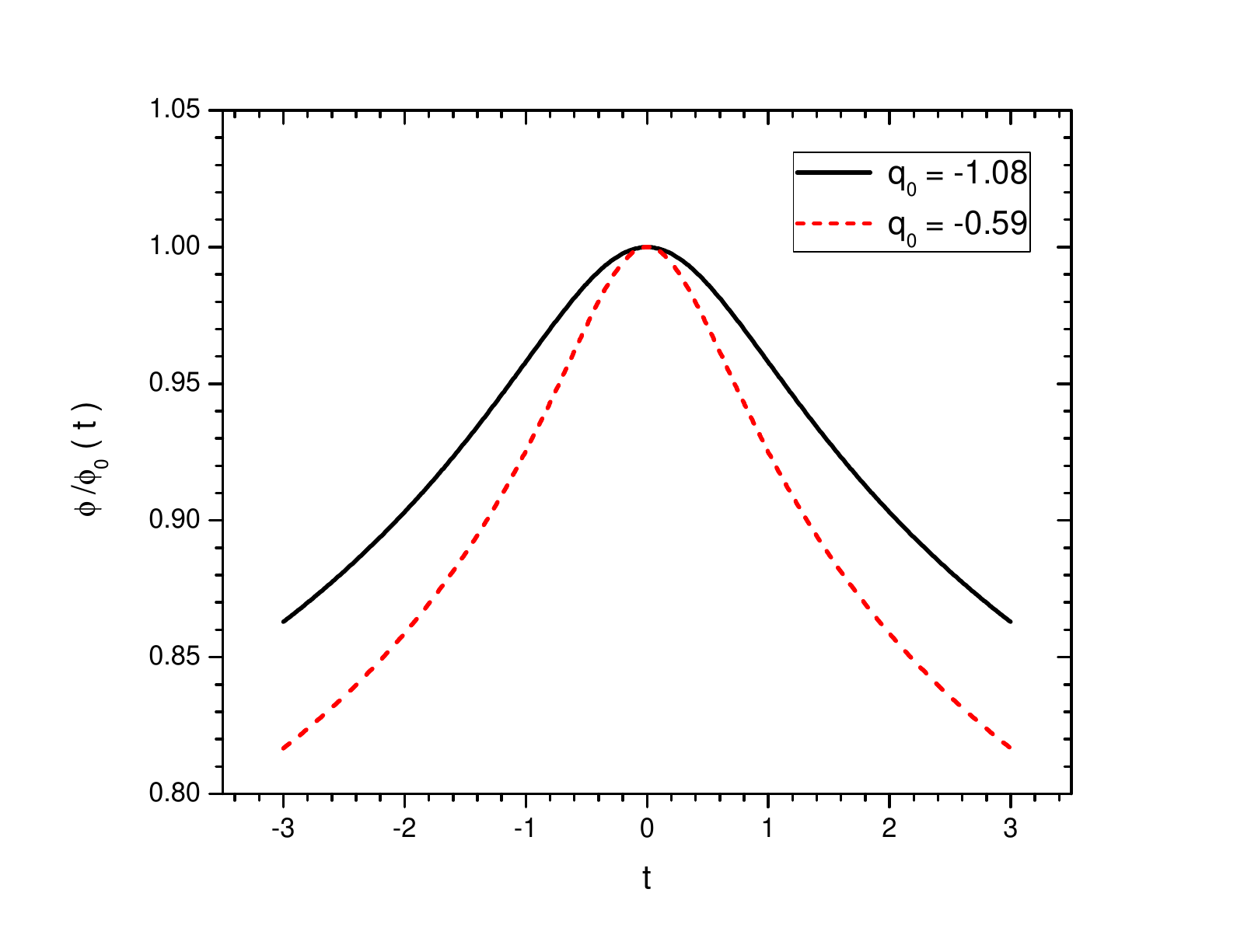}
\caption{Evolution of the scalar field.}\label{FIG.2}
\end{figure}

The evolution of the scalar field is shown in FIG.\ref{FIG.2}.  It is worth to mention here that, the scalar field and the scalar field potential depend on the choice of the parameter $m$. In general, the scalar field decreases with the growth of the scale factor and therefore, $m$ should be a negative quantity. In order to plot the figure, we chose a representative value for $m$ as $m=-0.1$. Also, we chose $\phi_0=1$. The scalar field smoothly decreases from a peak at the bouncing epoch to lower values in both the sides of the bouncing point. The peak is the same for both the choices of $q_0$. The curve of the scalar field with higher value of $q_0$ decreases very fast as compared to that with a lower value of $q_0$. 

\begin{figure}[H]
\centering
\minipage{0.50\textwidth}
\includegraphics[width=\textwidth]{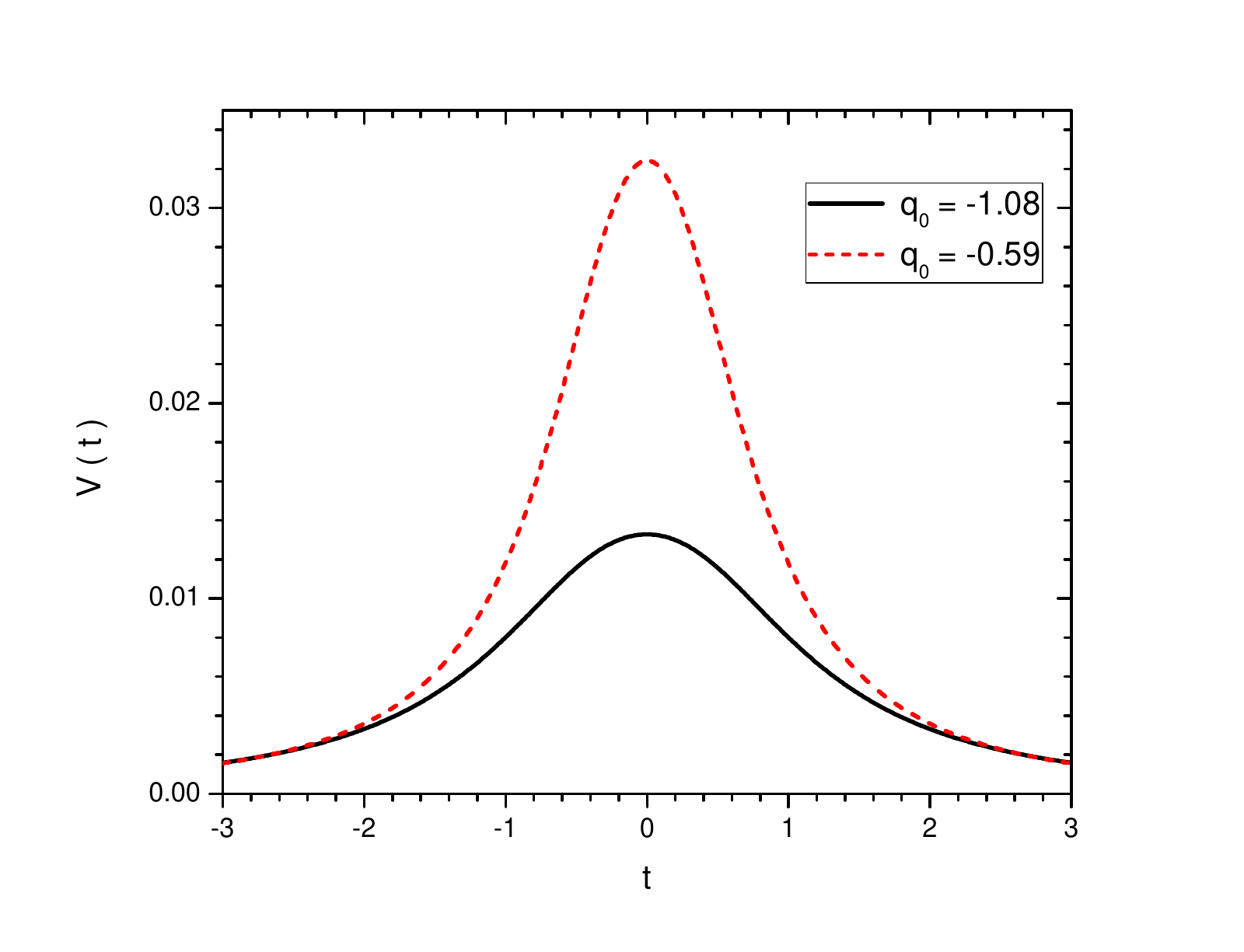}
\endminipage\hfill
\minipage{0.50\textwidth}
\includegraphics[width=\textwidth]{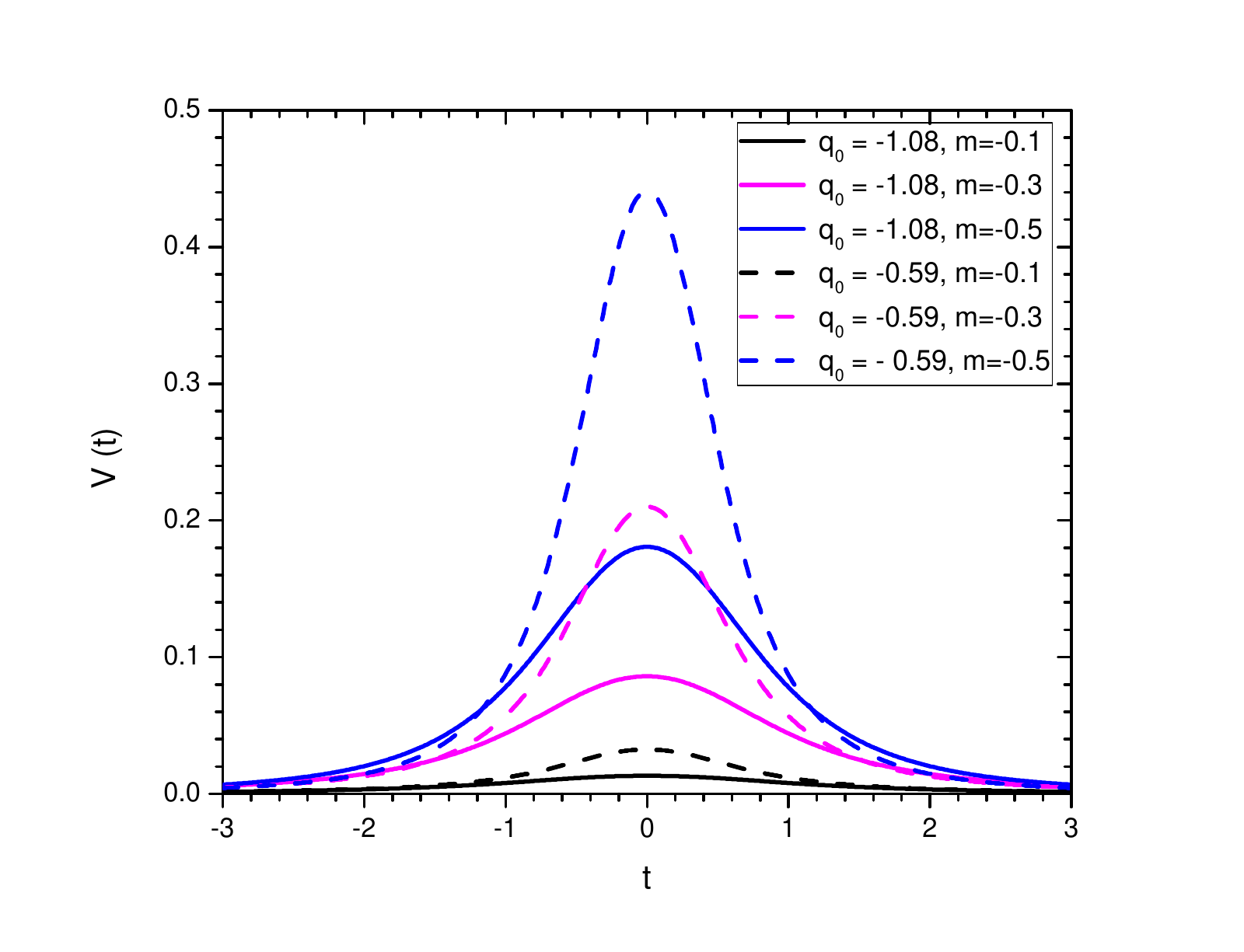}
\endminipage\hfill
\minipage{0.50\textwidth}
\includegraphics[width=\textwidth]{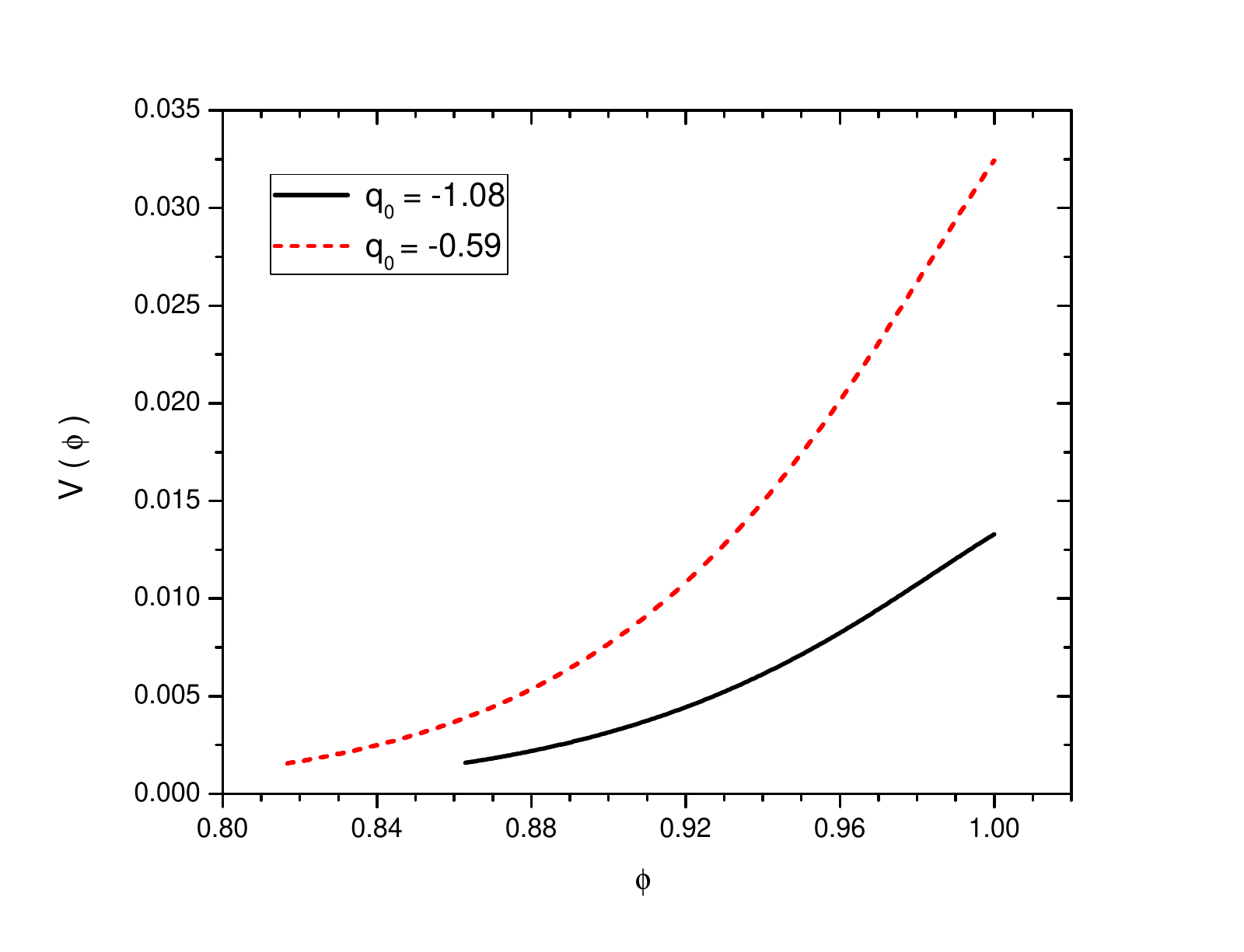}
\endminipage\hfill
\caption{(a) Top panel: Scalar field potential as a function of cosmic time. (c) Middle panel: Scalar field potential for different choices of the deceleration parameter and the exponent $m$. (c) Bottom panel: Scalar field potential as a function of the scalar field. }\label{FIG.3}
\end{figure}

The scalar field potential as reconstructed from the extended matter bounce scenario is shown  both as a function of cosmic time and scalar field in FIG. \ref{FIG.3}. The present bouncing scenario requires a repulsive scalar field potential upon which the scalar field rolls down to both the sides from the bouncing epoch. Obviously, the peak of the scalar field potential occurs at the bouncing epoch and it depends on the choice of $q_0$ and $m$. For a given value of $m$, one may note from the middle panel figure that, higher the value of $q_0$, higher is the peak value of the potential. However, for a given value of $q_0$, lower the value of $m$, higher is the peak of the potential. In the bottom panel of FIG. \ref{FIG.3}, the scalar field dependence of the potential is shown.  The scalar field potential shows an increasing tendency with the scalar field.

\section{Cosmological Implications}\label{SECIV}
It will be interesting to apply the reconstructed scalar field potential to study the dynamical features in some of the well known $f(T)$ models. For this purpose, we consider two functional forms of $G(T)$ that provide viable cosmological models.
\subsection{Model I}\label{sec:model_1}

Let us consider a specific choice
\begin{equation}\label{19}
    G(T)= \beta T \ln \left(\frac{T}{T_{0}}\right),
\end{equation}
where $\beta$ be a constant and $T_0$ be the value of $T$ at the initial epoch \cite{zhang2011notes}. The effective dark energy density and the effective dark energy pressure terms in Eqs.~\eqref{16}--\eqref{17} for this choice of the functional $G(T)$ become
\begin{eqnarray}
    \rho_{d}&=&\frac{\dot{\phi}^{2}}{2}+V(\phi)-6\beta H^{2} \ln \left(\frac{6 H^{2}}{T_{0}}\right)- 12 H^{2} \beta\,, \label{20} \\ 
    p_{d}&=&\frac{\dot{\phi}^{2}}{2}-V(\phi)+6\beta H^{2} \ln \left(\frac{6H^{2}}{T_{0}}\right)+ 12 H^{2} \beta 
    \nonumber\\
    & &+ 4 \dot{H} \left(\beta \ln \left(\frac{6H^{2}}{T_{0}} \right)+3 \beta \right). \label{21}
\end{eqnarray}

In terms of the parameter of the scale factor, the effective dark energy density may be expressed as
\begin{eqnarray}
    \rho_d &=  \frac{32}{9} m^2 \phi_0^2 \alpha \left[\frac{(1+\alpha t^2)^{n-1}}{n-1}-\frac{(1+\alpha t^2)^n}{n}  \right]\nonumber\\
    &- \frac{32}{3}\beta \alpha^2~t^2 (1+\alpha t^2)^{-2}\left[\ln \left(\frac{32\alpha^2~t^2}{3T_0(1+\alpha t^2)^2}\right)+2\right]
\end{eqnarray}
where $n=\frac{4m-3}{3}$.

The effective dark energy pressure now becomes
\begin{eqnarray}
    p_d &=& \frac{32}{3}\beta \alpha^2~t^2 (1+\alpha t^2)^{-2}\left[\ln \left(\frac{32\alpha^2~t^2}{3T_0(1+\alpha t^2)^2}\right)+2\right]\nonumber\\
    &+& \frac{16}{3}\alpha\beta\frac{1-\alpha t^2}{(1+\alpha t^2)^2}\left[\ln \left(\frac{32\alpha^2~t^2}{3T_0(1+\alpha t^2)^2}\right)+3\right].
\end{eqnarray}

From Eqs.~\eqref{20}--\eqref{21}, the equation of state parameter can be obtained as 
\begin{eqnarray}\label{26}
    \omega_{d}&= & \frac{\dot{\phi^{2}}-2 V(\phi)+12 H^{2} \beta \ln\left(6\frac{H^{2}}{T_{0}}\right)+24 H^{2} \beta}{\dot{\phi^{2}}+2 V(\phi)- 12 H^{2} \beta \ln\left (6\frac{H^{2}}{T_{0}}\right)-24 H^{2} \beta }\nonumber \\
   &  &+\frac{8 \dot{H} \left(\beta \ln (6\frac{H^{2}}{T_{0}})+3 \beta \right)}{\dot{\phi^{2}}+2 V(\phi)- 12 H^{2} \beta \ln\left (6\frac{H^{2}}{T_{0}}\right)-24 H^{2} \beta}\,.
\end{eqnarray}

\begin{figure}[ht!]
\centering
\includegraphics[scale=0.38]{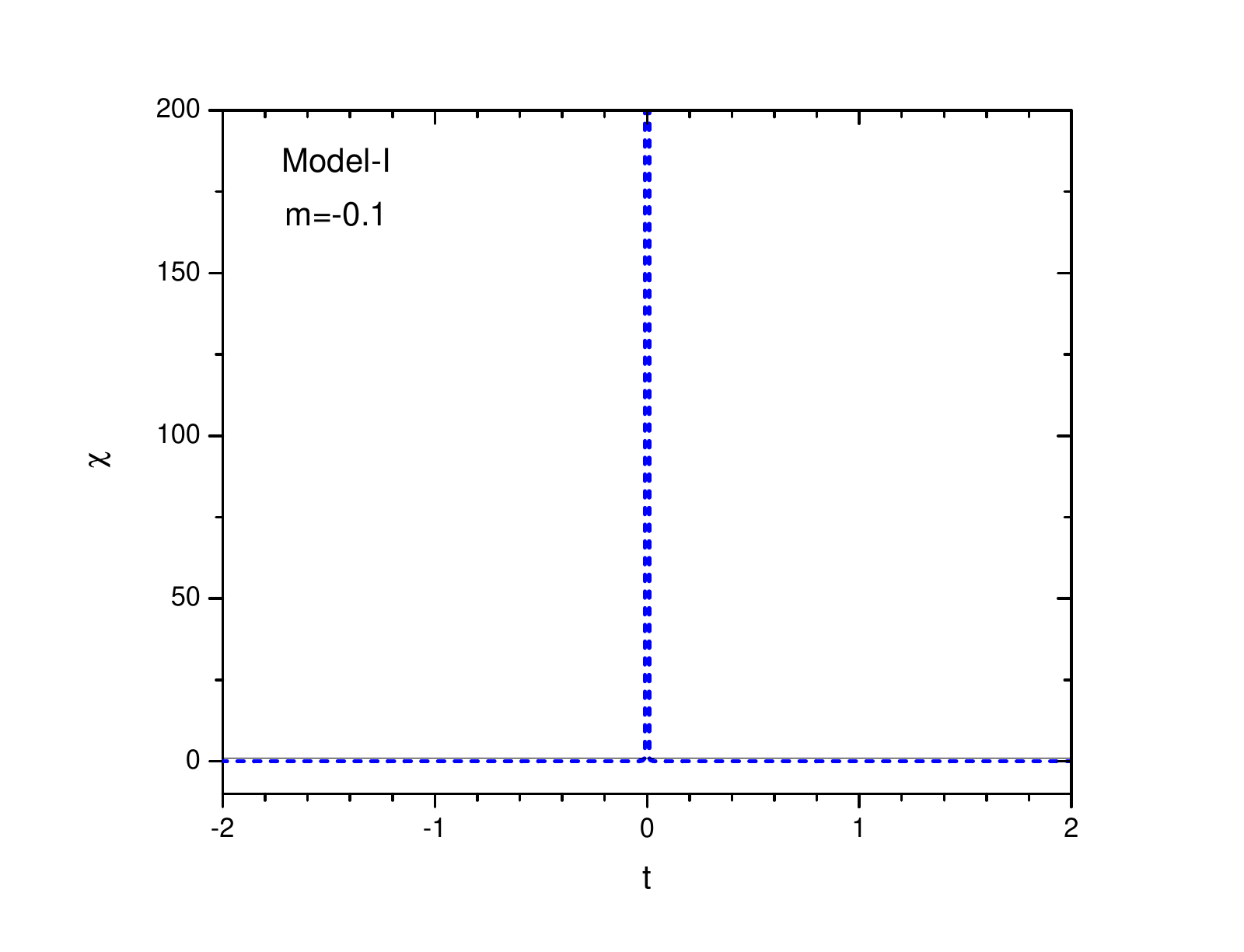}
\caption{Validity of the model through a check with \eqref{11}.}\label{FIGvalidityI}
\end{figure}

It is necessary that, the present model with the ansatz \eqref{22} and \eqref{19}, the Friedman equations \eqref{11} and\eqref{12} should be satisfied at least near the bouncing epoch. In view of this, we define a quantity
\begin{equation}
    \chi= \frac{\frac{\dot{\phi}^2}{2}+V(\phi)-2TG_{T}+G(T)+\rho_{m}+\rho_{r} }{3H^2}.
\end{equation}
The satisfaction of the Friedman equation for the present model requires that the quantity $\chi$ should be nearly equal to 1. In order to check the departure of the present model, we plot the quantity $\chi$ as a function of time $t$ in FIG. \ref{FIGvalidityI}. It is clear from the figure that, the bouncing ansatz \eqref{22} for the present gravity model satisfies Eq.\eqref{11}. At the bounce, the Hubble parameter vanishes and therefore the behaviour of $\chi$ shoots up and may not be conclusive.

\begin{figure}[ht!]
\centering
\includegraphics[scale=0.38]{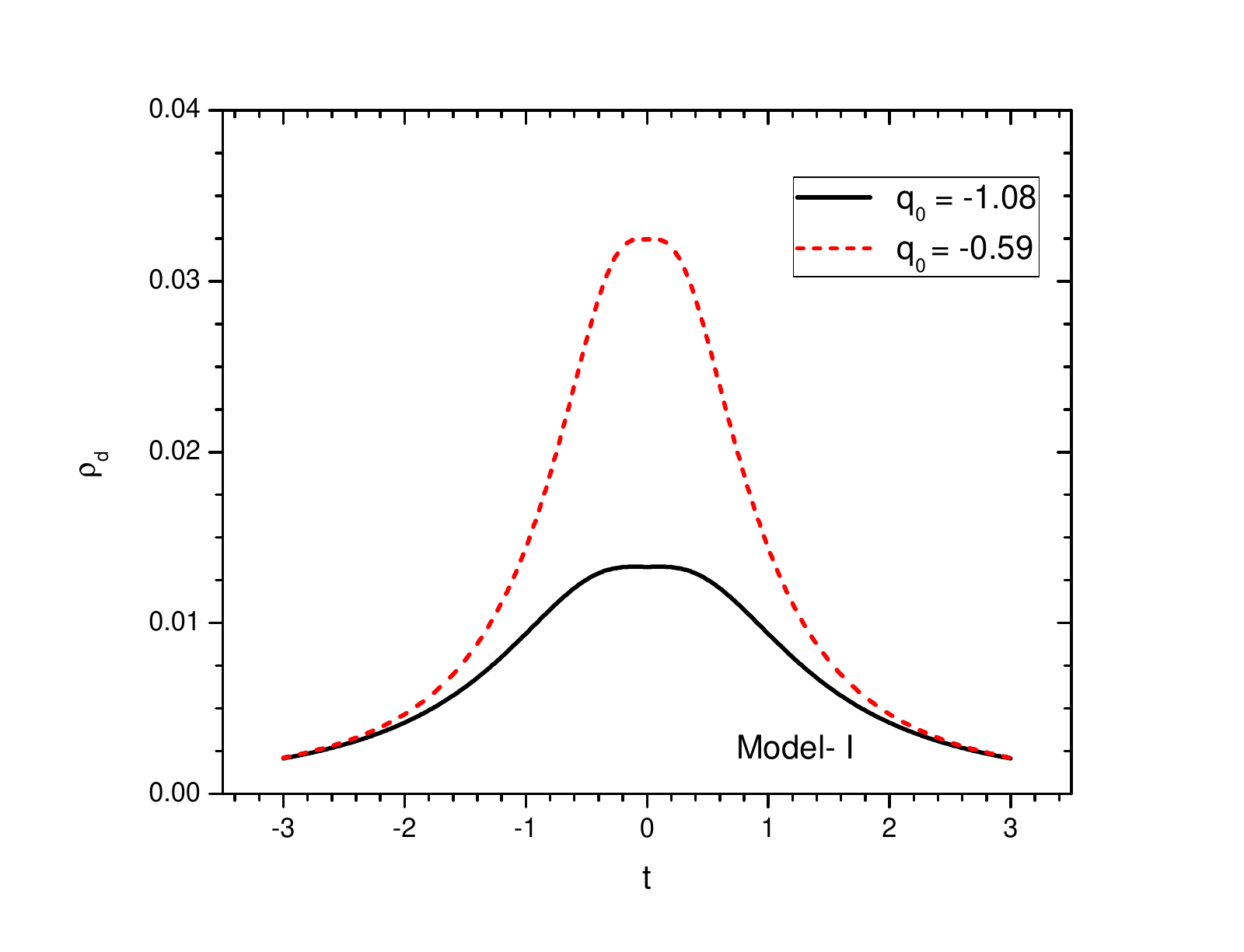}
\caption{Dark energy density for Model-I for two values of the deceleration parameter at the present epoch.}\label{FIG.4}
\end{figure}

\begin{figure}[ht!]
\centering
\includegraphics[scale=0.38]{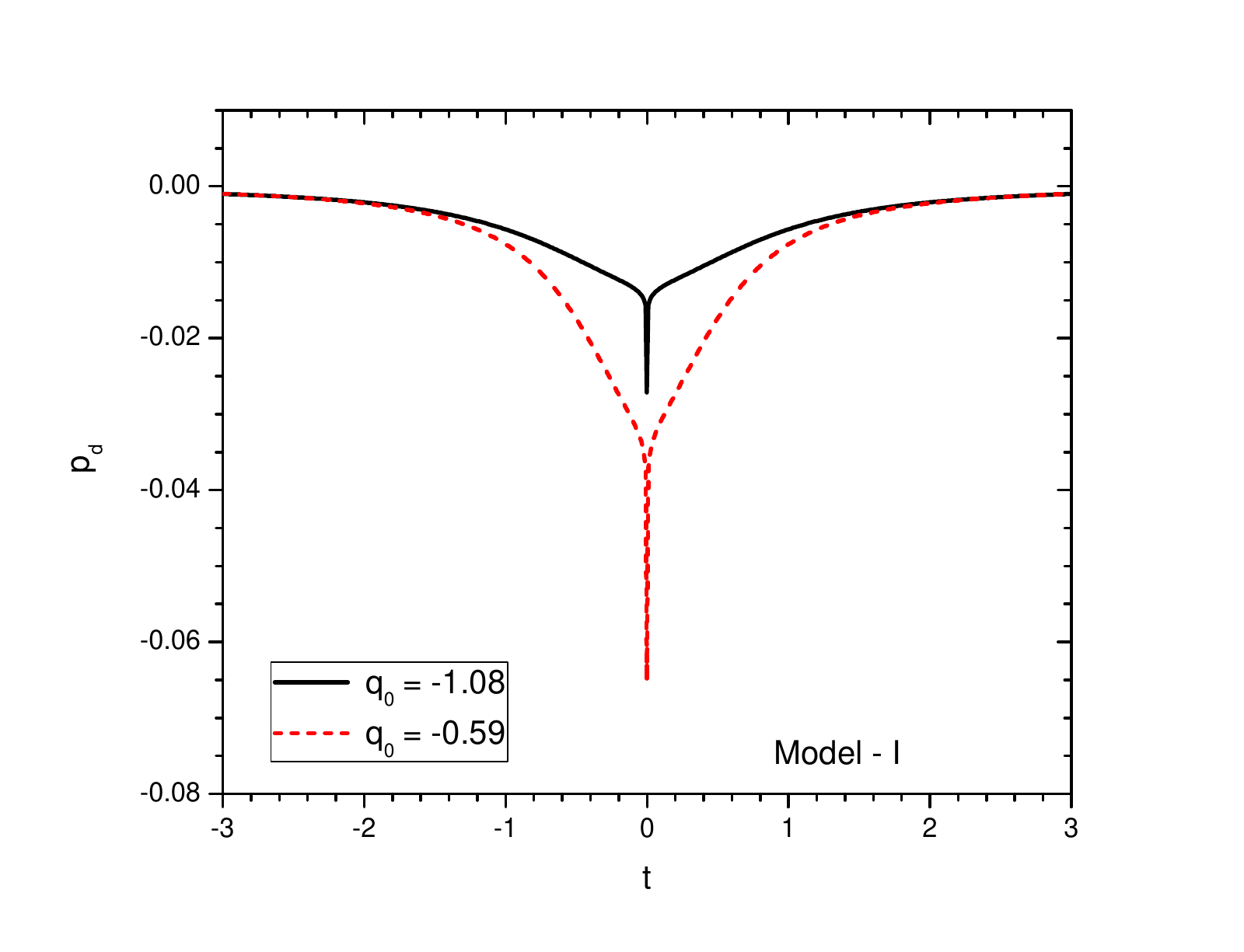}
\caption{Dark energy pressure for Model-I.}\label{FIG.5}
\end{figure}

\begin{figure}[ht!]
\centering
\includegraphics[scale=0.38]{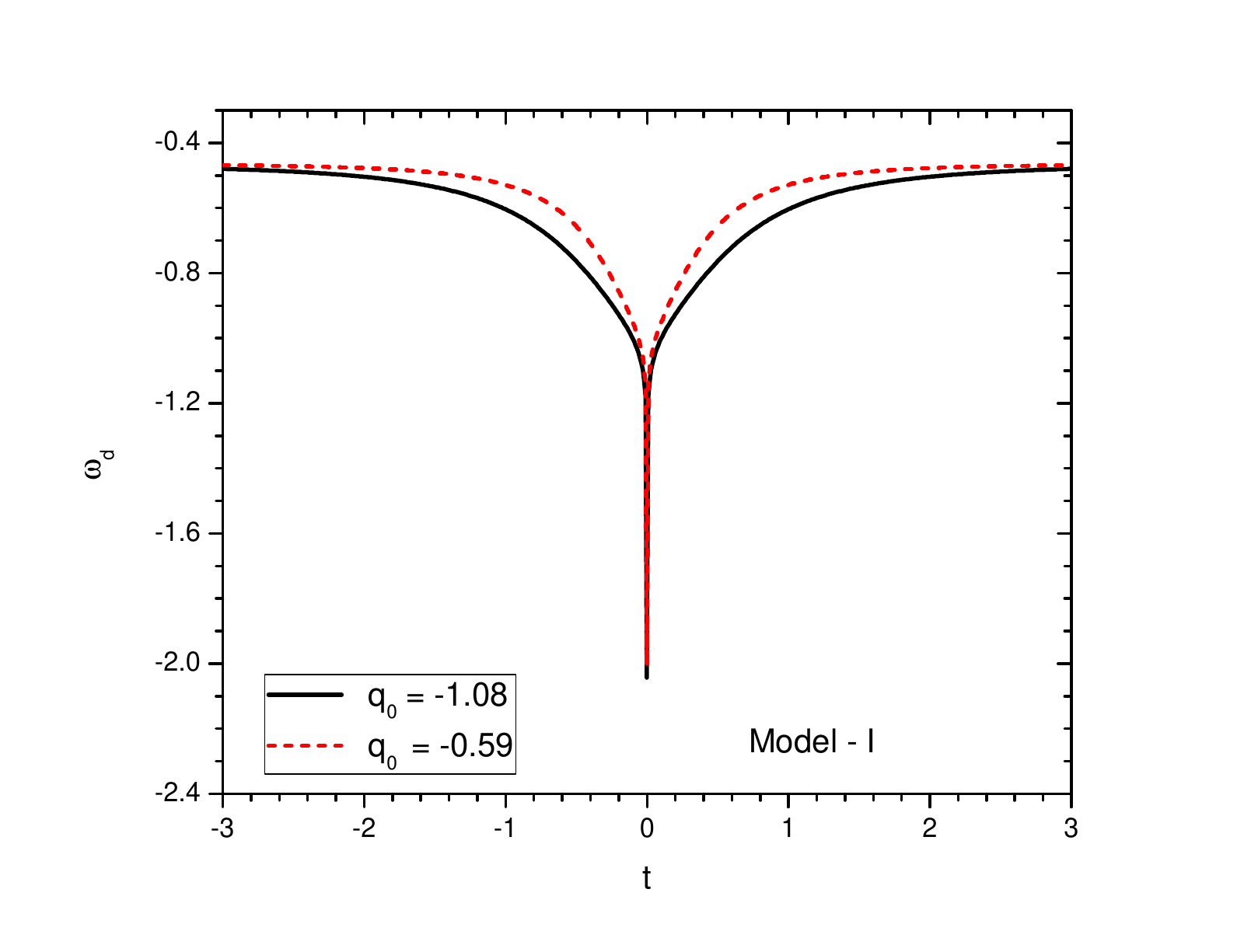}
\caption{Evolution of the dark energy EoS parameter for Model-I.}\label{FIG.6}
\end{figure}

The behaviour of the dark energy density, dark energy pressure and dark energy EoS parameter are shown in FIGs. \ref{FIG.4},\ref{FIG.5} and \ref{FIG.6} respectively. The parameters of the present model are fixed in such a manner that, we get positive dark energy density in both the positive and negative time zones. The dark energy density shows a peak at the bouncing epoch, the magnitude of which depends on the choice of the value of the deceleration parameter at the present time. The peak of the dark energy density is higher for lower magnitude of $q_0$. As time extends away from the bounce, the dark energy density falls off smoothly. On the otherhand, the dark energy pressure becomes a negative quantity showing a sharp ditch at the bouncing epoch. It increases from the sharp minimum in both the sides of the bounce. The depth of the sharp minimum is more for higher magnitude of $q_0$. For this model, the dark energy EoS parameter is also negative for the the time zone considered in the work. Just like the dark energy pressure, it shows a sharp minimum at the bouncing epoch but its evolutionary trajectory has a higher slope for lower magnitude of $q_0$ unlike the dark energy pressure. 

\subsection{Model II}\label{sec:model_2}

In this case, we consider the form of $G(T)$ as, 
\begin{equation}
    G(T)= T+\gamma T^{2},\label{33}
\end{equation}
where $\gamma$ is a constant \cite{Fortes:2021ibz}, which is a small generalization beyond TEGR.  For $\gamma=0$ the model reduces to the TEGR model. This may also be regarded as a truncated form of the previous model.

 The dark energy density of this model is obtained as
    \begin{eqnarray}
        \rho_{d}&=&\frac{\dot{\phi^{2}}}{2}+V(\phi)-6H^2(1+6 \gamma H^2)\,,\nonumber\\ 
       &=& \frac{8}{9} \phi_0^2 \left[\alpha~ t^2\left(\alpha m^2-1\right)+\frac{6}{2m-3}\right] (1+\alpha t^2)^ \frac{4m-6}{3} \nonumber\\
        &-& \frac{32}{3(4m-3)} \alpha \phi^2   (1+\alpha t^2)^\frac{4m-3}{3}\nonumber\\
        &-&\frac{32}{3}\alpha^2 t^2 (1+\alpha t^2)^{-2}\left[1+\frac{32}{3}\alpha^2\gamma^2 t^2 (1+\alpha t^2)^{-2}\right].
\end{eqnarray}

Similarly, we obtain the dark energy pressure of this model as
\begin{eqnarray}
    p_{d} &=& \frac{\dot{\phi^{2}}}{2}-V(\phi)+6H^2(1+18\gamma H^2)+4 \dot{H}(1+6 T \gamma)\,,\nonumber\\
    &=& m^2\phi^2~H^2+4\dot{H}\left(1+36\gamma~H^2\right)-\rho_d\,.
\end{eqnarray}

The Equation of State parameter (EoS) for the present model becomes
\begin{eqnarray}
    \omega_{d} &=& -1 + \left [\frac{p_{d}+\rho_d}{\rho_{d}}\right]\nonumber\\
    &=& -1+\frac{m^2\phi^2~H^2+4\dot{H}\left(1+36\gamma~H^2\right)}{\frac{\dot{\phi^{2}}}{2}+V(\phi)-6H^2(1+6 \gamma H^2)}.
\end{eqnarray}

\begin{figure}[ht!]
\centering
\includegraphics[scale=0.38]{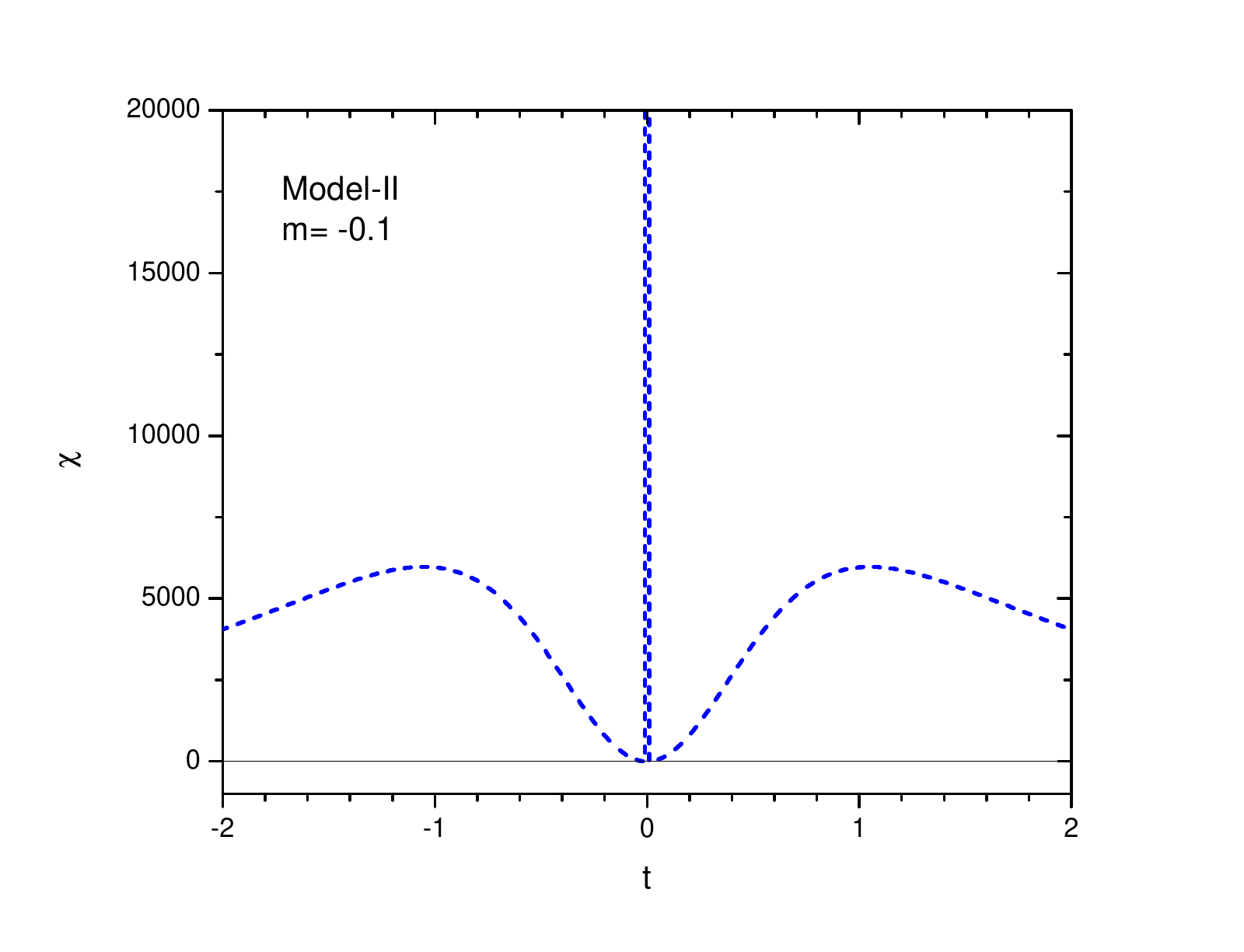}
\caption{Validity of the model through a check with \eqref{11}.}\label{FIGvalidityII}
\end{figure}

In FIG. \ref{FIGvalidityII}, the plot for $\chi$ is shown for the Model-II. According to the behaviour as displayed in the figure, $\chi$ shows a departure from unity in most of the time zone. However, in a very short time frame close to the bouncing epoch i.e. in the time region $-0.0125\leq t \leq 0.0125$, the ansatz \eqref{22} satisfies \eqref{11}. In this case also, the quantity $\chi$ shoots up due to the vanishing nature of the Hubble parameter at bounce.

\begin{figure}[ht!]
\centering
\includegraphics[scale=0.38]{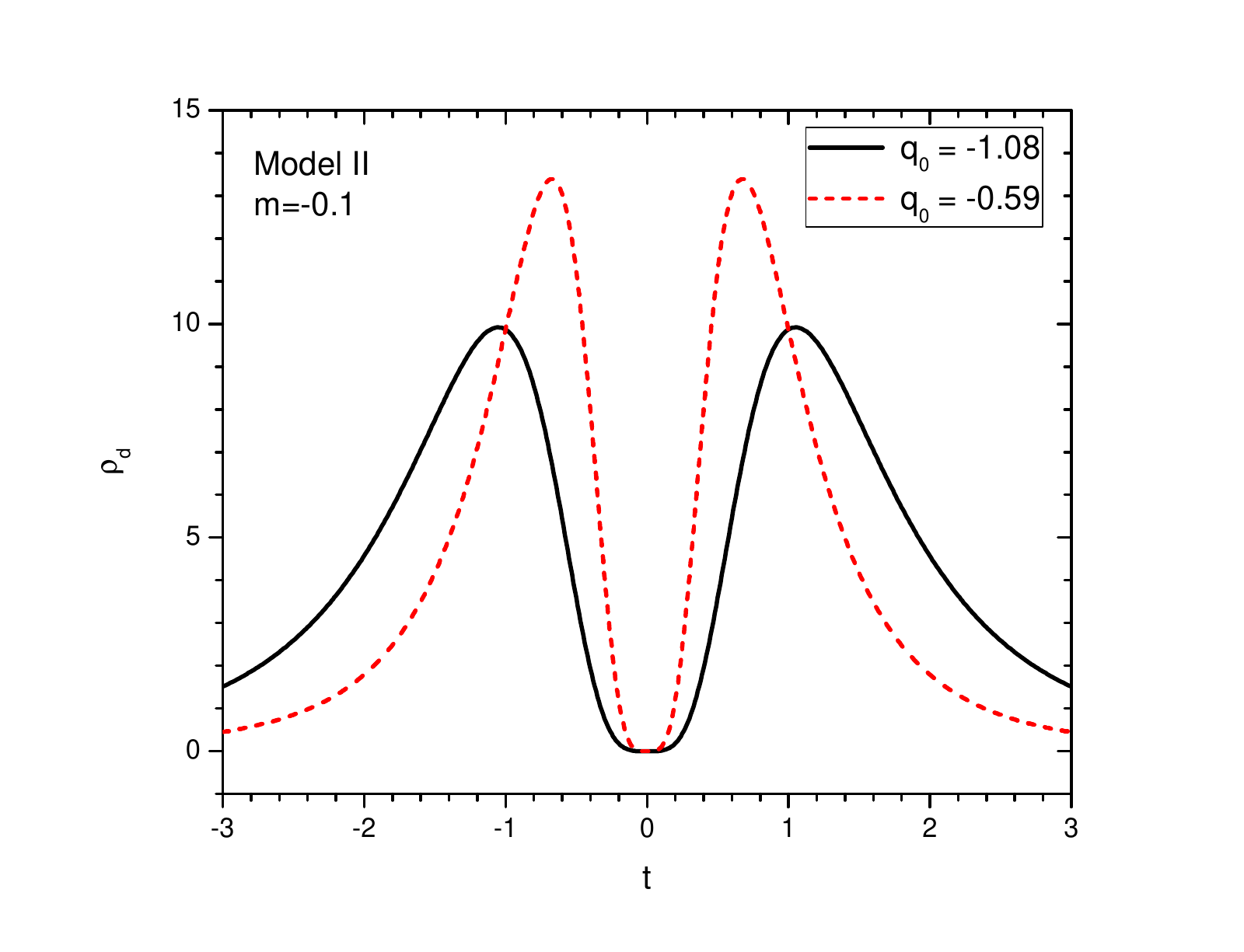}
\caption{Dark energy density for Model-II for two values of the deceleration parameter at the present epoch.}\label{FIG.7}
\end{figure}

\begin{figure}[ht!]
\centering
\includegraphics[scale=0.38]{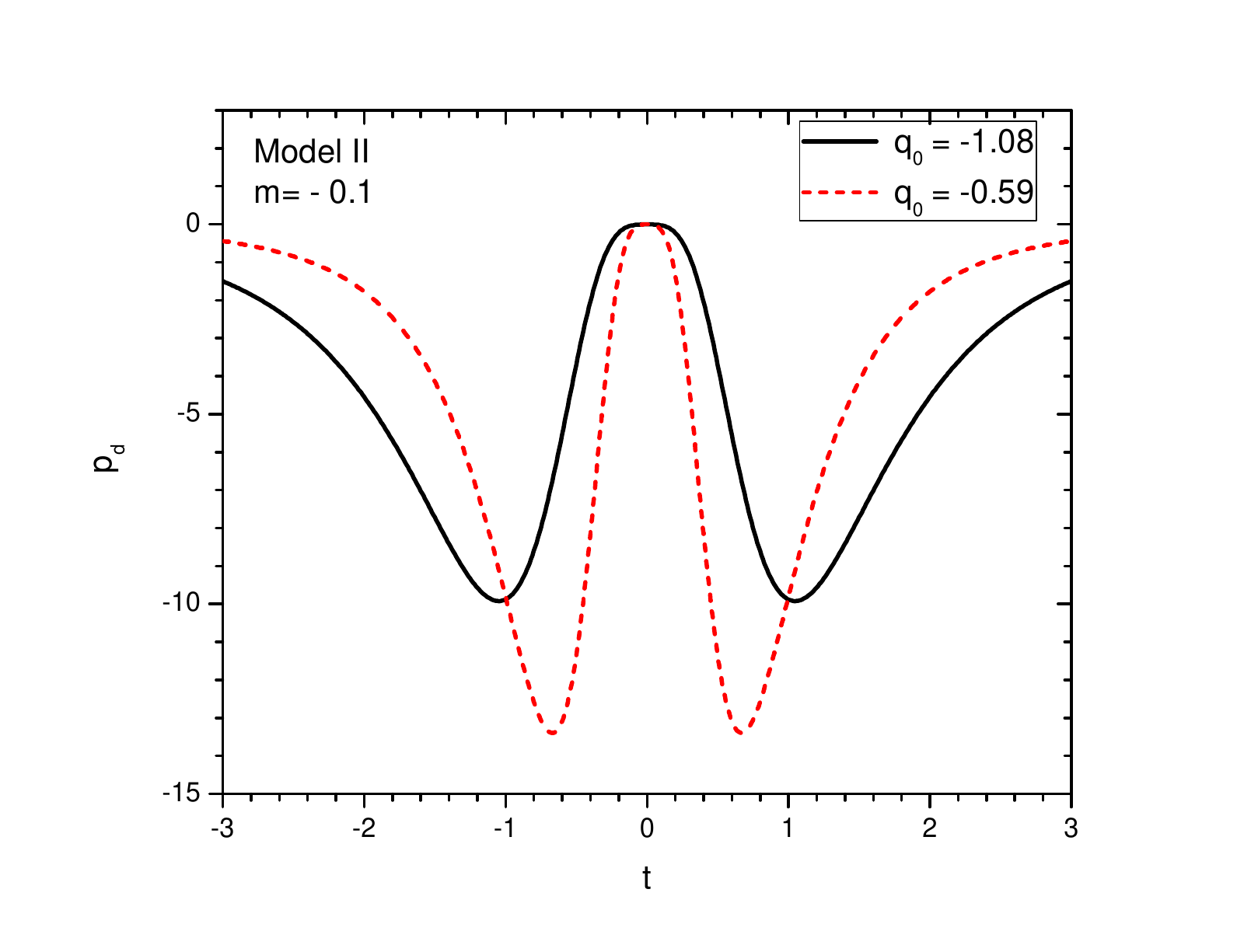}
\caption{Dark energy pressure for Model-II.}\label{FIG.8}
\end{figure}

\begin{figure}[ht!]
\centering
\includegraphics[scale=0.38]{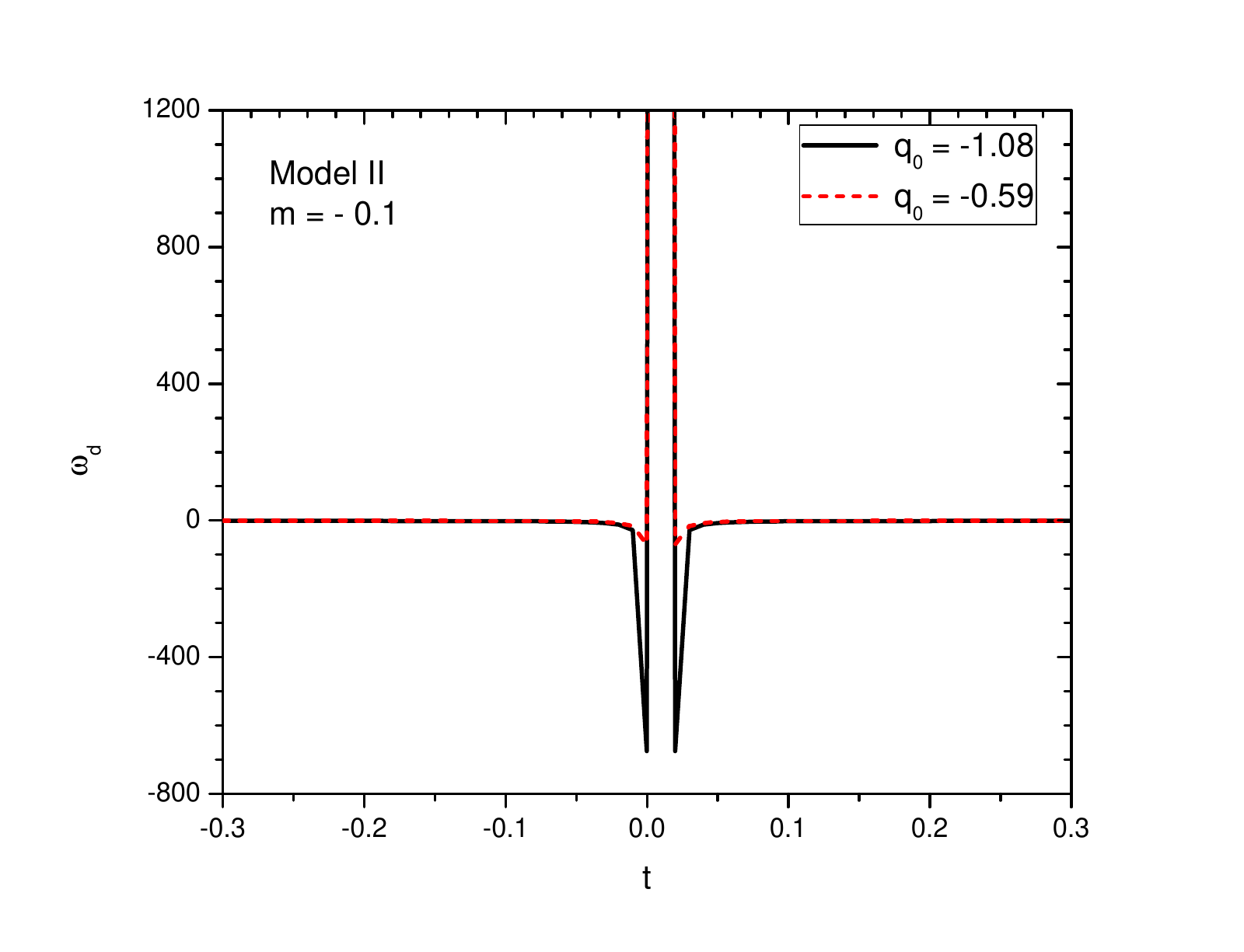}
\caption{Evolution of the dark energy EoS parameter for Model-II.}\label{FIG.9}
\end{figure}

The evolutionary aspects of the dark energy density, dark energy pressure and the dark energy EoS parameter are shown in FIGs. \ref{FIG.7}, \ref{FIG.8} and \ref{FIG.9} 
 respectively. We consider the same representative values of the parameters $\alpha, m$ and $\phi_0$ as in the previous model. The value of $\gamma$ is chosen to be $-0.03$ so that the dark energy density remains positive for the time zone considered in the work. Also, while choosing the value of $\gamma$, we keep in view that, the dominant energy condition (DEC) should not be violated. Unlike the previous model, the dark energy density shows an interesting behaviour. It has two lobes one in the pre-bounce phase and the other at the post bounce phase. At the bouncing epoch, the energy density drops to very lower values. Considering a specific time zone, for example the positive time zone, the dark energy density is observed to increase initially from low values to have a maximum reaching at a relatively quicker time and then falls off as the time advances. For a given value of the parameter $m$, the height of the lobes depend on the choice of $q_0$. Higher is the magnitude of $q_0$, lower becomes the lobe height. Also, the choice of $q_0$ affects the occurrence time of the peak in the dark energy density. 

 The dark energy pressure [refer FIG.\ref{FIG.8}] is obtained to be negative in both the positive and negative time zones forming almost an inverted image of the dark energy density behaviour. It has two wells formed in both sides of the bouncing point. As evident, the well depth is decided by the choice of the present value of the deceleration parameter.

 The behaviour of the dark energy EoS parameter for the present model is quite interesting [refer FIG.\ref{FIG.9}]. Its value is close to zero for most of the time zone except a visible feature in a time period ranging from some recent past to recent future relative to the bouncing epoch. In the post bounce phase, the dark energy EoS parameter decreases to more negative values after crossing the phantom divide and suddenly becomes positive just prior to the bounce. At the bounce, its value drops a little bit and increases again after bouncing occurs. At a small period after bounce, $\omega_d$ suddenly drops to have large negative values which subsequently raises to gain small negative values.
 
 \section{Energy Conditions}\label{SECV}
 In GR, Einstein's field equations address the causal metric and geodesic structure of the space-time, so the energy momentum tensor has to satisfy some conditions. For a space-time with signature $(-,+,+,+)$, we can take the time like vector $u^{i}$ to be normalized as $u_{i}u^{i}=-1$ and the future directed null vector $k^{i}$ as $k^{i}k_{i}=0$. The energy conditions may be defined as the contractions of time like or null vector fields with respect to Einstein tensor and the energy momentum tensor from the matter side of Einstein's field equations. Four energy conditions may be obtained such as:  
\begin{itemize}
\item At each point of the space time, the energy momentum tensor should satisfy, $T_{ij}u^{i}u^{j}\geq0$: Weak Energy Condition (WEC). So, $\rho\geq0$, $\rho+p\geq0$.
\item   For the future directed null vector $k^{i}$, $T_{ij}u^{i}u^{j}\geq0$: Null Energy Condition (NEC). So, $\rho+p\geq0$. 
\item  The matter flows along time like or null line and with contracted energy momentum tensor, the quantity $-T_{i}^{j} u^{ij}$ becomes future directed time like or null like vector field: Dominant  Energy Condition (DEC). So, $\rho-p\geq0$.
\item $\left(T_{ij}-\frac{1}{2}T g_{ij}\right)u^{i}u^{j}\geq0$ says the gravity has to be attractive: Strong Energy Condition (SEC). So, $\rho+3p\geq0$.
\end{itemize}

The null energy condition is a fundamental concept in general relativity that states the energy density measured by any observer can not be negative. From the perspectives of bouncing cosmology, it is imperative that the cosmic fluid should violate the null energy condition. In the present work, we have segregated the dark energy density and the corresponding pressure from the bulk contribution and investigated their importance in studying the cosmic dynamics. In view of this, here we consider the energy conditions for the dark sector part of the field equations.

\begin{figure}[ht!]
\centering
\includegraphics[scale=0.32]{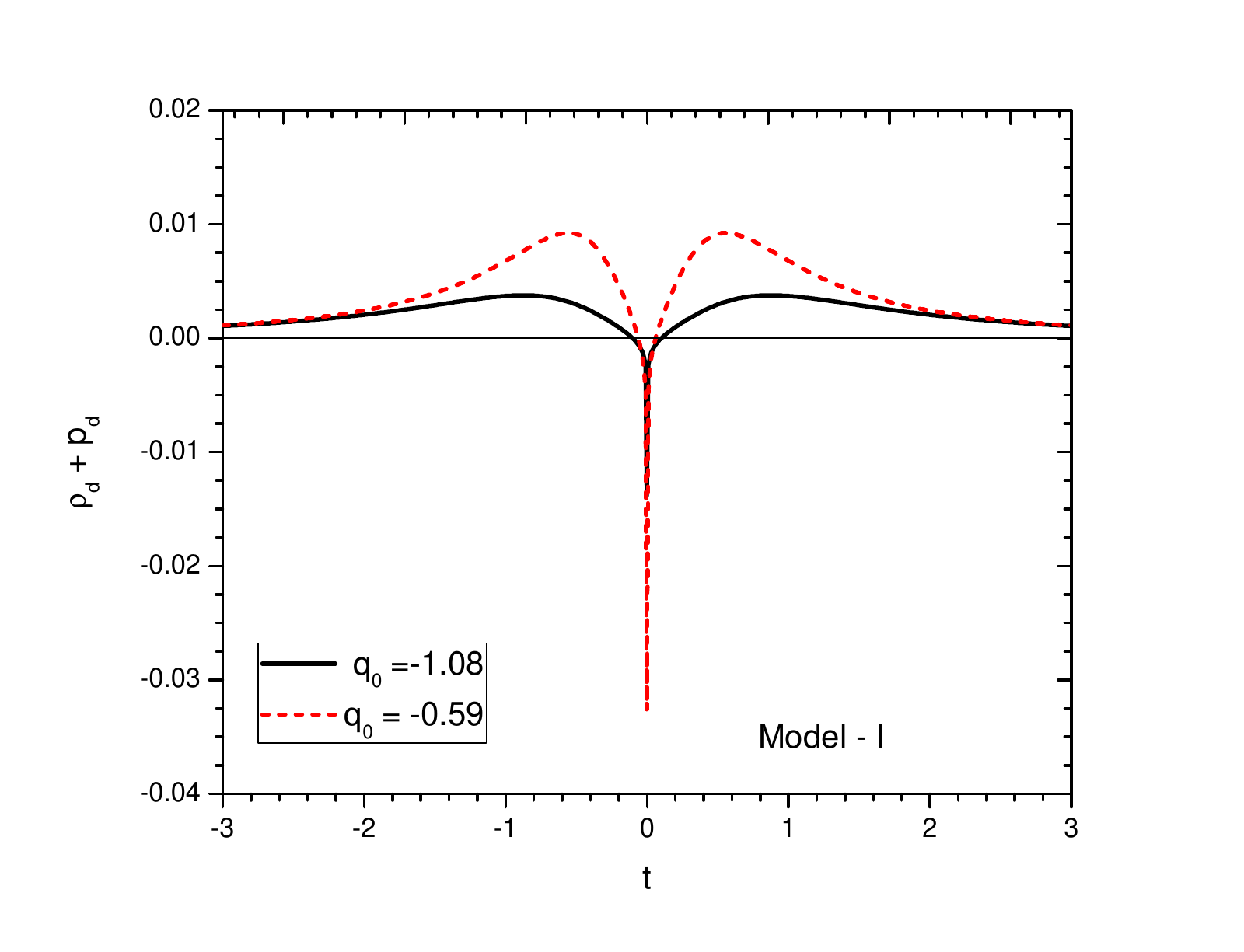}
\caption{Null energy condition for the Model-I.}\label{FIG.10}
\end{figure}

\begin{figure}[H]
\centering
\minipage{0.47\textwidth}
\includegraphics[width=\textwidth]{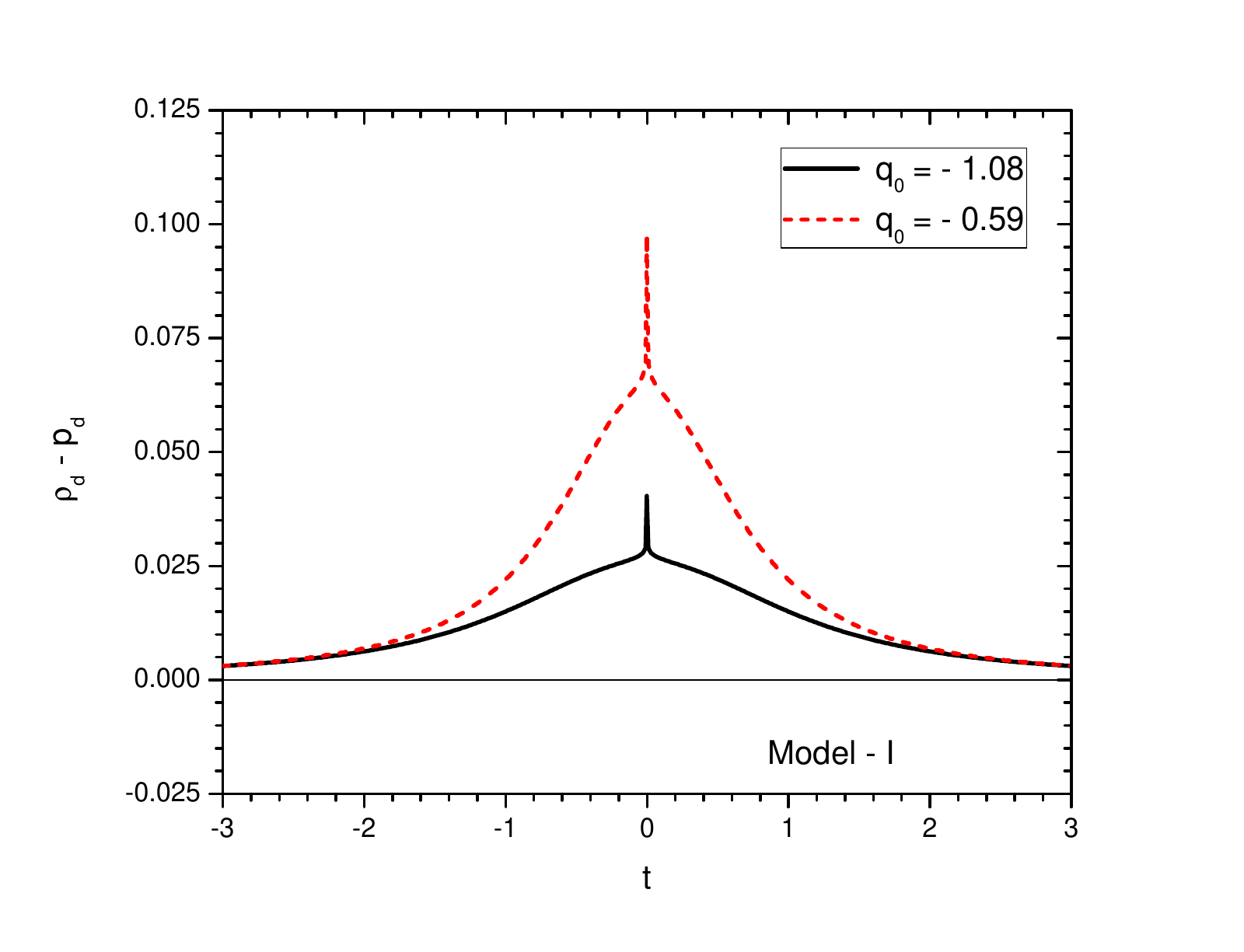}
\endminipage\hfill
\minipage{0.47\textwidth}
\includegraphics[width=\textwidth]{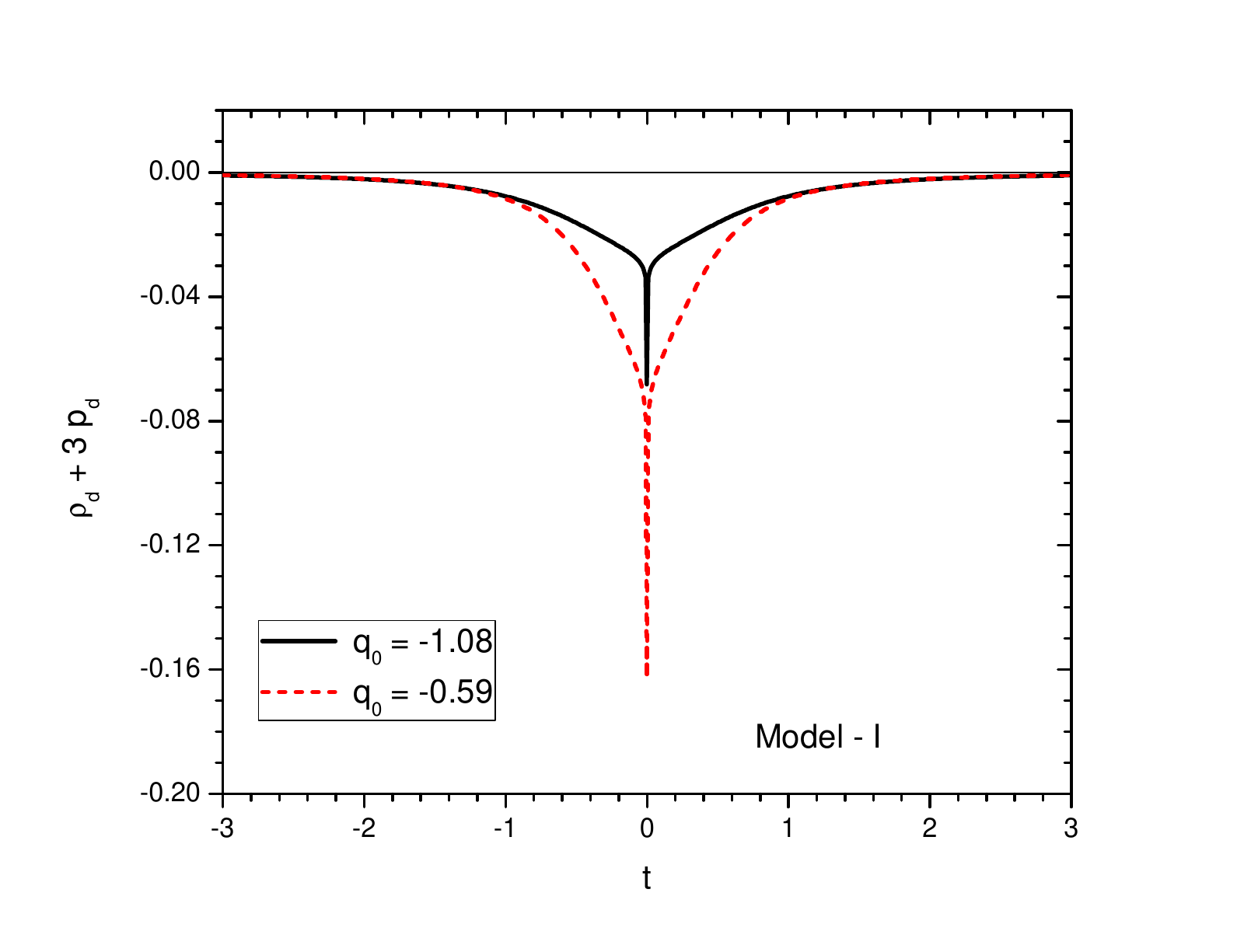}
\endminipage\hfill
\caption{(a) Top panel: DEC for Model-I. (b) Bottom panel: SEC for Model-I.}\label{FIG.11}
\end{figure}

\begin{figure}[ht!]
\centering
\includegraphics[scale=0.38]{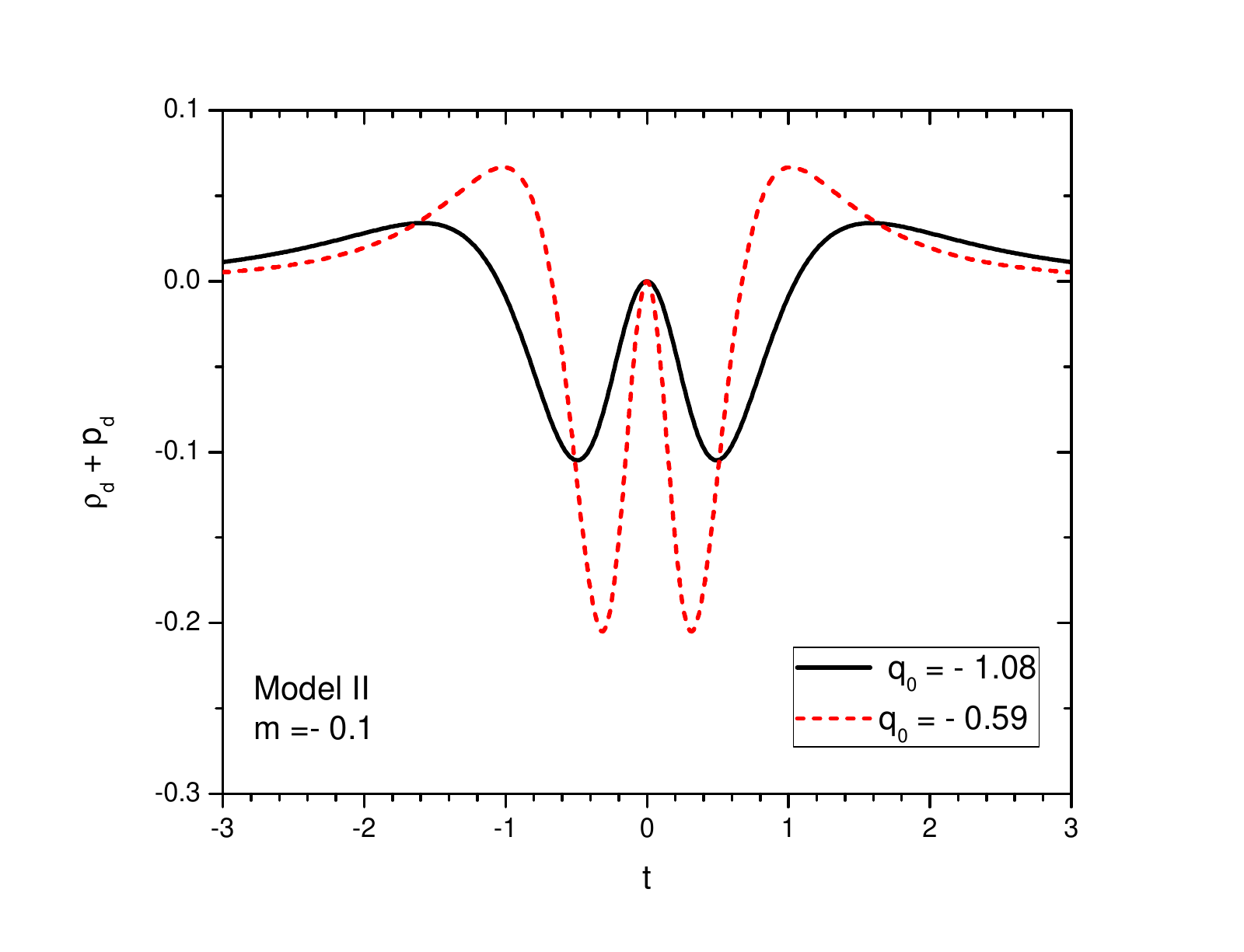}
\caption{Null energy condition for the Model-II.}\label{FIG.12}
\end{figure}

\begin{figure}[H]
\centering
\minipage{0.50\textwidth}
\includegraphics[width=\textwidth]{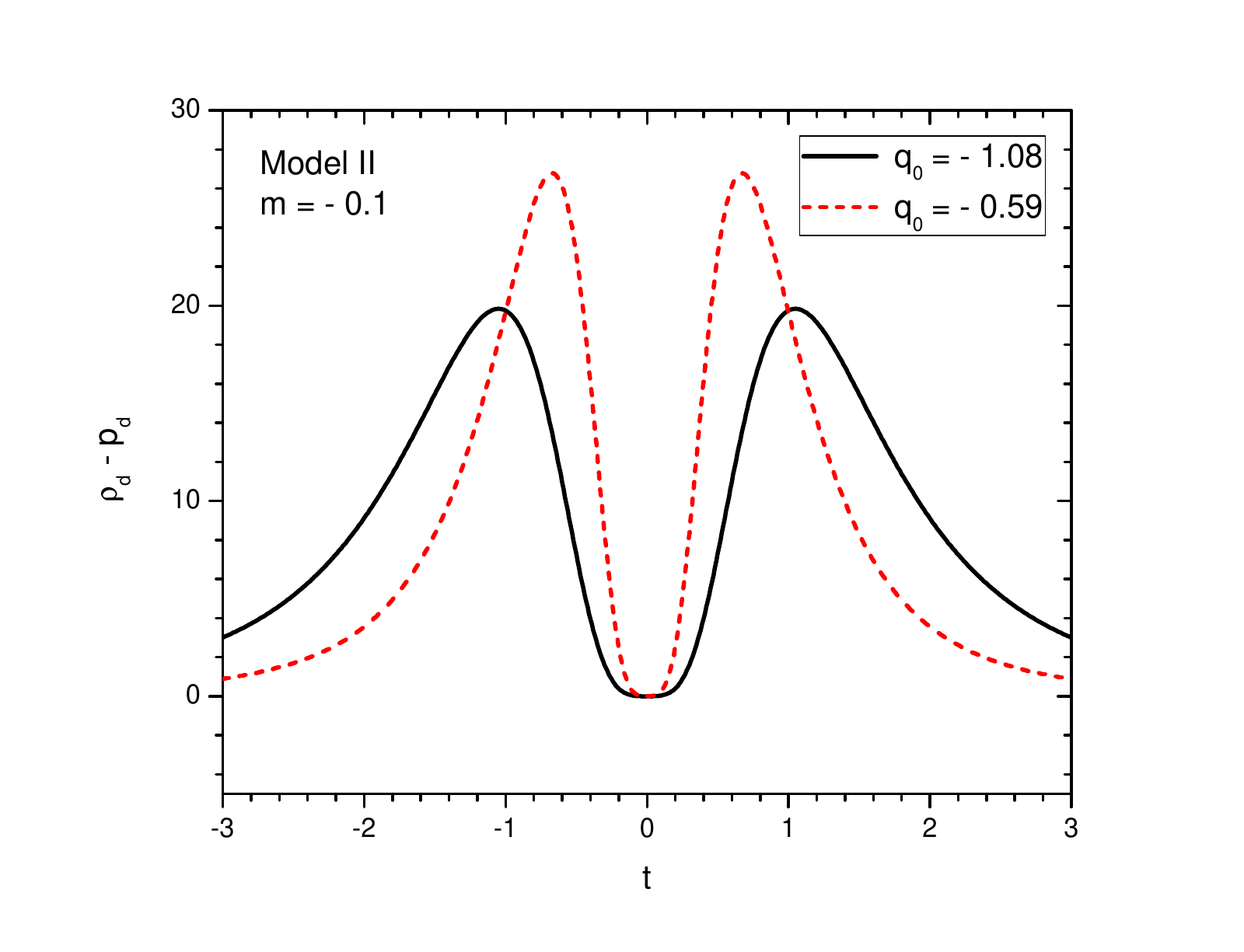}
\endminipage\hfill
\minipage{0.50\textwidth}
\includegraphics[width=\textwidth]{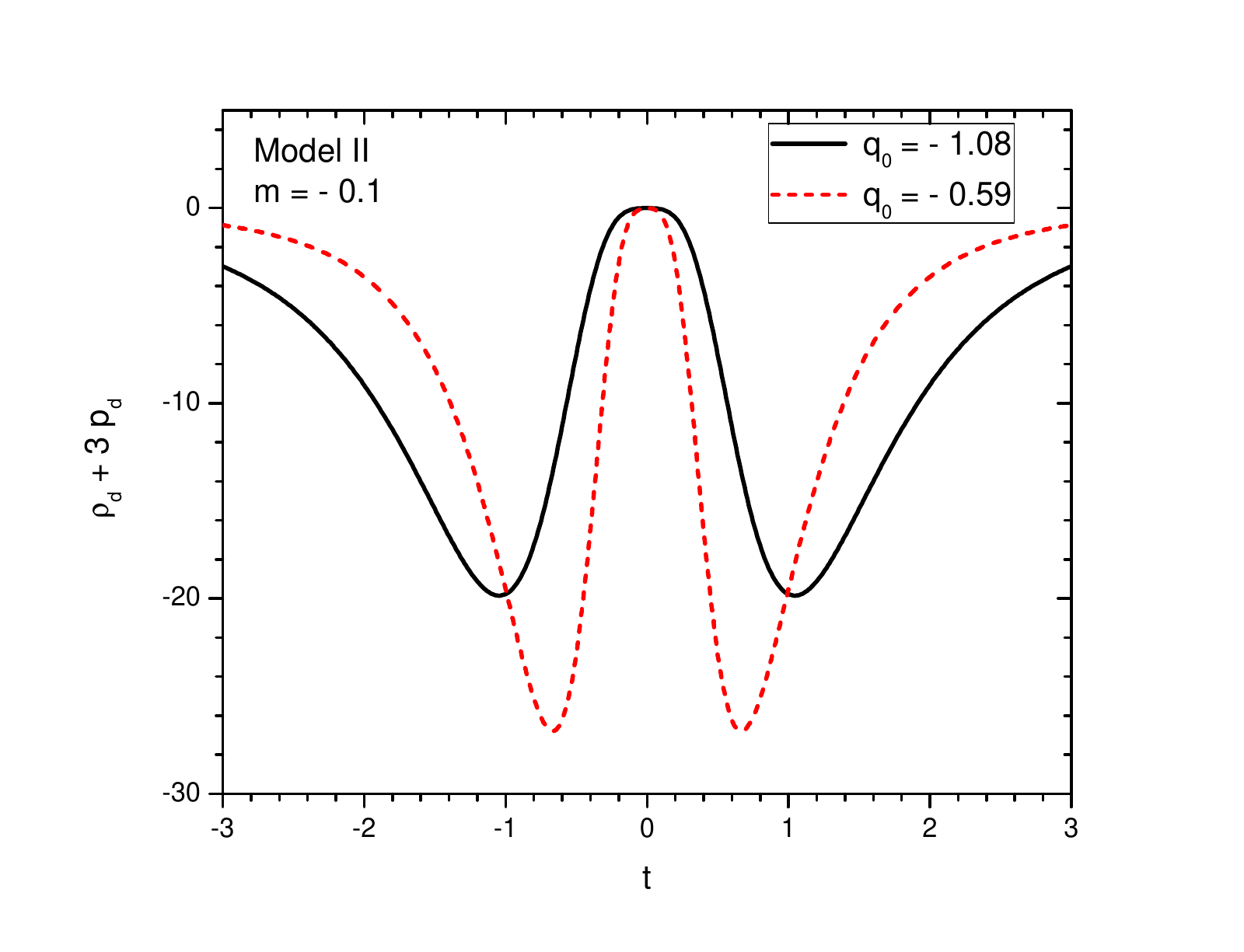}
\endminipage\hfill
\caption{(a) Top panel: DEC for Model-II. (b) Bottom panel: SEC for Model-II.}\label{FIG.13}
\end{figure}

The energy conditions for Model-I are plotted as function of time in FIGs.\ref{FIG.10} and \ref{FIG.11} and that for Model-II are shown in FIGs. \ref{FIG.12} and \ref{FIG.13}. In the case of Model-I, the NEC is satisfied in most of the time zone except near the bounce where the sum of the dark energy density and the dark pressure suddenly becomes negative and quickly after bounce, it becomes positive again. For a narrow time zone near the bouncing epoch, the NEC is violated ensuring the usual nature of a bouncing scenario. For this model, for whole of the time frame considered, the DEC is satisfied and the SEC is violated envisioning an accelerated cosmic expansion.

For Model-II, the NEC is violated for a narrow time zone near the bounce. However, the time period for which the NEC is violated is greater than that for Model-I. The DEC for this case is satisfied. On the otherhand, the SEC is violated throughout the cosmic time zone. However, the violation of SEC is not ensured for the whole time zone as there may be a possibility of SEC satisfaction or at least the sum  $\rho_d+p_d$ vanishes in a time zone close to the bounce for some choices of the model parameter $\gamma$ and the deceleration parameter at the present epoch. 

\section{Conclusion} \label{SECVI}
Different aspects of cosmological issues have been studied using the teleparallel theory of gravity $f(T)$ in recent times. This gravity theory is considered as one of the most suitable alternative to GR where we have used the geometrical aspects of torsion scalar instead of the usual Ricci scalar. A simple extension of $f(T)$ is the $f(T,\phi)$ gravity theory where the geometric part of the action contains a self interacting scalar field $\phi$ with a scalar potential $V(\phi)$. In usual scalar-tensor theories of gravity, the role of the scalar potential is quite significant. Also, in inflationary scenario or bouncing scenario the  scalar field potential  plays interesting roles to explain the occurrence of such scenarios. In view of this it is natural to consider a self interacting scalar field with a potential in the field equations of the teleparallel gravity theory. 

In the present work, we consider such a teleparallel theory along with the scalar field contribution and reconstructed the scalar field potential required to witness an extended matter bounce scenario. From an integration of the Klein-Gordon equation for the scalar field, the scalar field potential is reconstructed for the assumed bouncing cosmic behaviour. The reconstruction of the scalar field potential is found to depend upon the choice of the deceleration parameter at the present epoch. We consider two recently constrained values of the deceleration parameter at the present time. The scalar field is obtained to decrease smoothly from a peak occurring at the bouncing epoch. It is interesting to note that, the peak of the scalar field is independent of the choice of $q_0$. The extended matter bounce scenario considered within the framework of the given gravity theory requires a repulsive scalar field potential that rolls down to both the side of the peak. Considering the whole evolutionary period spanning from the pre-bounce phase to post-bounce phase,  the model requires a maximum scalar potential for the bouncing scenario to occur, the value of which depends on the choice of the present epoch value of the deceleration parameter. For lower magnitude of the deceleration parameter, the repulsion becomes stronger at the bouncing epoch. 

The cosmological implications of the reconstructed scalar field potential are investigated for two well known functions of the torsion scalar. For both the cases, the model parameters are fixed up in such a manner to retrieve the geometrical dark energy density to have positive values for the whole time zone considered in the work. The associated dark energy pressure and the dark energy EoS are obtained. For the first model, the dark energy density shows a peak at the bounce and falls off as we move away from the bounce, whereas the dark energy pressure shows a sharp ditch with negative values. Consequently, the dark energy EoS also shows a sharp ditch near the bounce with large negative values. On the otherhand, for the second model, the dark energy density has two lobes in both the sides of the bouncing epoch. The dark energy density  assumes lower values near the bounce. The dark pressure has two ditches in both the sides of the bouncing region, portraying a kind of mirror reflection of the corresponding dark energy density. The dark energy EoS parameter displays an interesting behaviour for Model-II, where it crosses the phantom divide in the pre-bounce phase to have large negative values and just before the bouncing epoch it suddenly becomes positive. A polar opposite kind of behaviour is displayed in the post-bounce phase.

The null energy condition is expected to be violated in bouncing scenario i.e. during the bouncing phase, the sum of the energy density and the pressure should remain positive. As a part of studying the cosmological implications with the reconstructed scalar field potential, the NEC is obtained for both the models. While for Model-I, the NEC is satisfied for most of the time zone, it is violated close to the bouncing epoch. On the other hand, for Model-II, the violation or satisfaction of the null energy condition depends on the choice of the deceleration parameter at the present epoch. While the NEC is satisfied at an epoch away from the bounce, the sum of the dark energy density and dark energy pressure assumes a much smaller value at the bounce.  Although, there have been some studies with the ansatz \eqref{33} in addressing different cosmological issues, in the present investigation of the quantum mechanically motivated matter bounce scenario, it shows some poor feature. In this sense, the Model-I may be having a priority over the other and may provide a viable cosmology with potential avenues for further investigation.

\section*{Acknowledgements} BM acknowledges SERB-DST for the Mathematical Research Impact Centric Support (MATRICS)[File No : MTR/2023/000371]. SKT and BM thank IUCAA, Pune (India) for providing support in the form of an academic visit during which a part of this work is accomplished. 

\section*{References}
\bibliographystyle{utphys}
\bibliography{references}

\end{document}